\documentclass[fleqn,usenatbib]{mnras}

\usepackage{newtxtext,newtxmath}
\usepackage{pdflscape}
\usepackage{threeparttable}
\usepackage{threeparttablex}
\RequirePackage{rotating}
\usepackage{rotating}

\usepackage[T1]{fontenc}
\usepackage{hyperref}
\DeclareRobustCommand{\VAN}[3]{#2}
\let\VANthebibliography\thebibliography
\def\thebibliography{\DeclareRobustCommand{\VAN}[3]{##3}\VANthebibliography}

\usepackage{graphicx}	
\usepackage{amsmath}	
\usepackage{subfigure}
\usepackage{float}
\usepackage{threeparttable}
\usepackage{threeparttablex}
\usepackage{xspace}

\graphicspath{{Images/}}
\usepackage[utf8]{inputenc}
\usepackage[normalem]{ulem}

\title{Detecting eclipsing double white dwarfs with electromagnetic and gravitational waves}

\author[H. M. Jin et al.]{
Hong-Ming Jin,$^{1}$
Bo Ma,$^{2,3}$
Yong Shao,$^{4}$
and Yan Wang$^{1}$\thanks{E-mail: ywang12@hust.edu.cn (YW)}
\\
$^{1}$MOE Key Laboratory of Fundamental Physical Quantities Measurements, PGMF, 
Department of Astronomy and School of Physics, \\
Huazhong University of Science and Technology, Wuhan 430074, P. R. China \\
$^{2}$School of Physics and Astronomy, Sun Yat-sen University, Zhuhai 519082, P. R. China\\
$^{3}$CSST science Center in the Great Bay Area, Zhuhai 519082, P. R. China\\
$^{4}$Department of Astronomy, Nanjing University, Nanjing 210023, P. R. China\\
}

\date{Accepted 2024 December 04. Received 2024 November 29; in original form 2024 June 12}

\pubyear{2022}

\begin{document}
\label{firstpage}
\pagerange{\pageref{firstpage}--\pageref{lastpage}}
\maketitle

\begin{abstract}

Galactic double white dwarfs are predominant sources of gravitational waves in the millihertz frequencies accessible to space-borne gravitational wave detectors. With advances in multi-messenger astronomy, an increasing number of double white dwarf systems will be discovered through both electromagnetic and gravitational wave observations. In this paper, we simulated two populations of double white dwarfs originating from different star formation histories (hereafter referred to as Model 1 and Model 2) using the binary population synthesis method. We predicted the number of double white dwarfs in our Galaxy detectable by TianQin and Laser Interferometer Space Antenna (LISA) individually, as well as through their joint observation. In addition, we performed an analysis to evaluate the accuracy of the parameter estimation using the Fisher information matrix. Furthermore, we predicted the number of detached eclipsing double white dwarfs detectable by Gaia and the Vera C. Rubin Observatory (VRO). Our study found that over the nominal mission durations, TianQin, LISA, and their joint observation can detect at least five thousand and potentially several tens of thousands of double white dwarfs with signal-to-noise ratios greater than 7. Gaia and VRO are expected to detect at least several dozen and up to several hundred eclipsing double white dwarfs with orbital periods less than 30 hours. We also found that several dozen eclipsing double white dwarfs can be detected jointly through electromagnetic and gravitational wave observations.

\end{abstract}




\section{Introduction}

White dwarfs (WDs) are the final evolutionary stage of main sequence stars with masses less than $8-10~{ M_{\sun}}$ \citep{Iben_Jr__1997, 2006MNRAS.369..383D}. Based on our current understanding of stellar evolution and initial mass function, it is estimated that more than 95\% of main sequence stars in our Galaxy will evolve into WDs~\citep{2019BAAS...51c.123B,2021MNRAS.504.2707G}. These stars can no longer produce energy through nuclear fusion reactions, and as a result, they will gradually cool down over time by radiating their stored thermal energy, eventually becoming fainter. The evolution of WDs is reasonably well understood, with further details available in previous reviews~\citep{doi:10.1146/annurev.aa.28.090190.001035, Fontaine_2001, 2010A&ARv..18..471A}. For a discussion of recent advances and challenges in observations and theoretical modeling, see~\citep{2022PhR...988....1S,2024NewAR..9901705T,2024Ap&SS.369...43B}.

A significant fraction of stars in our Galaxy reside in binary systems~\citep{2012Sci...337..444S, doi:10.1146/annurev-astro-081710-102602, 10.1093/mnras/stab323}. Observation shows that the binary fraction is greater than 0.5 for stars with masses $>1~{ M_{\sun}}$~\citep[see][Appendix A]{2013A&A...552A..69V}. \citet{2017A&A...602A..16T} simulated a sample of WDs within 20~pc of Earth using a population synthesis method and predicted that optically resolvable (angular separation of orbit larger than a critical value) double white dwarfs (DWDs) constitute $9-14\%$ of the population. 
Previous simulations using binary population synthesis have suggested that there are about $10^8$ DWDs in our Galaxy \citep{2001A&A...375..890N, 2010A&A...521A..85Y, Breivik_2020}. 
With the advances of multi-messenger astronomy, a wealth of observations on WD binaries will be pivotal in exploring the evolution of binary stars~\citep{postnov2014evolution}, refining the parameters in binary population synthesis models, such as the efficiency parameter of common envelope evolution~\citep{2000A&A...360.1043D,2022ApJ...937L..42H, 2024arXiv240406845C}. In addition, these observations contribute to the exploration of the mechanisms underlying Type Ia supernovae~\citep{2012A&A...546A..70T} and to the study of the structure of our Galaxy~\citep{10.1111/j.1365-2966.2011.18976.x, 2019MNRAS.483.5518K}. 

Observing WDs is challenging due to their inherently low luminosities. However, the total number of WD candidates has steadily risen with the ongoing progress of large-scale automated surveys and the accumulation of data. The latest WD catalog from Gaia Early Data Release 3 (EDR3) contains about 359,000 high-confidence WD candidates~\citep{2021MNRAS.508.3877G}. 
Through a spectroscopic survey focusing on $<0.3~{M_{\sun}}$ He-core WDs, the Extremely Low Mass (ELM) survey has discovered 98 detached DWDs~\citep{2020ApJ...889...49B}, and more recently an additional 34 ELM WD binaries have been discovered in the southern sky~\citep{2020ApJ...894...53K,2023ApJ...950..141K}.Follow-up spectroscopy and high-speed photometry observations have confirmed a growing number of DWDs~\citep{2019Natur.571..528B, 2020ApJ...905...32B, 2020ApJ...905L...7B, 2020MNRAS.494L..91C, 10.1093/mnras/stab1318}. 

Eclipsing binary systems provide the most robust means of determining the parameters of both stellar components.  
Based on Gaia's WD catalog and data from the Zwicky Transient Facility (ZTF), 17 new eclipsing binaries have been identified \citep{2022MNRAS.509.4171K}, among which the majority are the WD binaries with a main-sequence companion, with only two possibly containing an extremely low-mass WD companion. Utilizing the Gaia EDR3 catalog and ZTF light curve data, \citet{2023ApJS..264...39R} identified a sample of 429 close WD binary candidates through a systematic search for short-period binaries. \citet{2017PASP..129f5003W} quantitatively validated the detection capability of VRO~\citep{2009arXiv0912.0201L} for eclipsing binaries by simulating a pseudo-VRO sample of binary light curves and subsequently recovering their periods from the light curves. The results indicate that 71$\%$ of the periods could be successfully recovered.
As the data from Gaia and ZTF continue to accumulate and with the future operation of the 10-year multi-purpose optical survey of VRO, an increasing number of DWDs will be discovered. 

On the other hand, in the era of gravitational wave (GW) astronomy, DWDs are one of the most promising sources for proposed space-borne laser interferometers, such as the Laser Interferometer Space Antenna \citep[{LISA};][]{2017arXiv170200786A}, TianQin \citep{TianQin2016}, Taiji \citep{10.1093/nsr/nwx116}, ASTROD-GW \citep{1998grco.conf..309N}, and gLISA \citep{Tinto2015gLISA}. 
Tens of thousands of DWDs in our Galaxy could be detected by these GW detectors ~\citep{2001A&A...365..491N, 2011CQGra..28i4019M, 2010ApJ...717.1006R, 10.1093/mnras/stx1285, 2019MNRAS.490.5888L, 2020PhRvD.102f3021H}. 
Binary systems that have been confirmed by EM observations and are expected to be detected by future GW observations are defined as verification binaries (VBs). 
To date, several tens of VBs have been discovered and their properties have been well characterized~\citep{2018MNRAS.480..302K, 2024ApJ...963..100K, 2020ApJ...905...32B, 2024MNRAS.532.2534M}. 

While more than one hundred DWD systems have been discovered so far by electromagnetic (EM) observations, there is a noticeable selection bias. This bias arises from the fact that DWDs with higher luminosities and closer distances to the observer are more likely to be detected by EM observations. However, to study the population characteristics of DWDs in our Galaxy, we need a more comprehensive sample. Binary population synthesis (BPS) is the conventional method for studying the evolution of binary populations and has wide applications in astrophysics. To date, a variety of BPS codes have been developed, such as SeBa \citep{1996A&A...309..179P,2001A&A...365..491N}, Yunnan Model \citep{1998MNRAS.296.1019H,2002MNRAS.336..449H,2003MNRAS.341..669H, 2002MNRAS.334..883Z,2004A&A...415..117Z}, BSE \citep{2002MNRAS.329..897H}, COSMIC \citep{2020ApJ...898...71B}, and COMPAS \citep{2022ApJS..258...34R}. Binary population synthesis produces statistical samples of certain types of stars for subsequent astrophysical investigations. For example, it has been used to explore the number of detectable DWDs in our Galaxy for LISA~\citep{2017A&A...602A..16T, 2020A&A...638A.153K}. \citet{2020PhRvD.102f3021H} investigated the ability of TianQin to detect DWDs in our Galaxy using a synthetic catalog from ~\citet{2012A&A...546A..70T,2017A&A...602A..16T}, which shows that about $8.7\times10^3$ DWDs can be detected over the 5-year nominal mission duration of TianQin. 
The formation and evolution of ultra-compact X-ray binaries (UCXBs) in the Galactic bulge has been explored using SeBa simulations by \citet{2013A&A...552A..69V}. \cite{2021ApJ...919..128R} investigated the potential of LISA to detect of GWs from inspiraling binaries in common envelope phases using COSMIC. \citet{10.1093/mnras/staa002} used a double neutron star (DNS) population generated by COMPAS to estimate the properties of DNS systems with cumulative SNRs greater than 8 during the 4-year LISA mission. Using a modified BSE code, \citet{Shao2021} predicted that LISA could detect dozens of GW sources with stellar-mass black holes in our Galaxy. 

The structure of this paper is as follows. In Section~\ref{sec:pop}, we introduce the population synthesis method used to generate DWD populations in our Galaxy. 
In Section~\ref{sec:emobs}, we simulate the light curves of eclipsing DWDs and investigate their detectability in the photometric surveys of Gaia and VRO. 
In Section~\ref{sec:gwobs}, we estimate the number of DWDs that are detectable by LISA, TianQin, and their joint observations. 
In addition, we evaluate the accuracy of the parameter estimation for these DWDs.
In Section~\ref{sec:joint}, we discuss the results of the joint observations of DWDs with GW and EM observations. The paper is concluded in Section~\ref{sec:sum}. 

\section{Population synthesis of DWDs}\label{sec:pop}

In this section, we discuss the settings of initial parameters, the choices of binary evolution models, and the Galactic structure when conducting binary population synthesis simulations for the DWDs in our Galaxy.

\subsection{Initial parameters }
\label{sec:maths} 
 
We initiate a population of zero-age main-sequence (ZAMS) stars by employing the probability distribution functions for their initial parameters. Subsequently, we evolve this initial population using a specific evolutionary model which leads to the formation of a sample of DWDs. 
The initial values for the parameters of the binaries are derived by sampling the distribution functions related to the initial metallicity ($Z$), primary mass ($m$), binary mass ratio ($q$), orbital separation ($a$), eccentricity ($e$), and orbital inclination angle ($\iota$). 
In our simulations, the initial mass function follows~\citet{1993MNRAS.262..545K} with a mass range of $0.08-150~M_{\sun}$, 
which provides a universally applicable model based on Galactic field data, reflects observed power-law index changes at critical masses ($0.08~M_{\sun}$ and $0.5~M_{\sun}$), and effectively quantifies observational uncertainties such as Poisson noise and dynamical evolution~\citep{2001MNRAS.322..231K}; The binary mass ratio $q$ adopts the uniform distribution between $[0,1]$ and this flat mass ratio distribution is a good first-order approximation when compared to observations~\citep{2013ARA&A..51..269D};
The orbital separation (in unit of solar radius) is sampled from the log-uniform distribution with $\log(a)$ between $[0,6]$~\citep{1983ARA&A..21..343A}, which is one of the most commonly used assumptions for the orbital semimajor axis distribution and is consistent with observations for solar mass and intermediate mass stars~\citep{2013ARA&A..51..269D}; 
and the eccentricity adopts the thermal distribution with $e$ between $[0,1]$\citep{1975MNRAS.173..729H}, which reflects the expectation from energy equipartition. In addition, because DWDs quickly circularize orbits in their early evolution, the initial eccentricity distribution has a limited effect on the simulation results.
We use a constant binary fraction of 0.5 and a constant metallicity of 0.02. 
Our initial parameters are the same as those in \citet{Breivik_2020}, but differ from the mass range ($0.95-10~M_{\sun}$) for the initial mass function used in \citet{2019MNRAS.483.5518K}. 

To investigate the impact of star formation history on binary system evolution, we simulate two populations of DWDs that share the same initial parameters described in the previous paragraph except for their star formation histories. 
In Model 1, we adopt a burst of star formation 13.7 Gyr ago~\citep{2010A&A...521A..85Y}, which can serve as a useful initial choice. Other stellar populations can then be obtained from this choice through convolution with specific star formation histories. In Model 2, we follow the star formation history in \citet{2022ApJ...937..118W}, which assumes a constant star formation rate over 10 Gyr for the thin disk, a 1 Gyr burst of star formation 11 Gyr ago for the thick disk, and a 1 Gyr burst of 10 Gyr ago for the bulge. 
This can be taken as an average of the star formation rate over certain cosmological epoch and is usually appropriate for low-redshift spiral galaxies~\citep{2014A&A...571A..72B}.

Due to limited computational resources, we only simulate a subset with a fixed number of DWDs. This sample will be scaled up to represent the entire population with  $N_{\rm Galaxy}$ DWDs, and  
\begin{equation}
    N_{\rm Galaxy} = N_{\rm fixed}\frac{M_{\rm Galaxy}}{M_{\rm fixed}} \,.
\end{equation}
Here, $N_{\rm fixed}$ denotes the number of DWDs in the fixed subset, $M_{\rm fixed}$ denotes the mass of the ZAMS stars used to produce this fixed subset, and $M_{\rm Galaxy}$ represents the mass of the different structures of our Galaxy. $N_{\rm Galaxy}$ is the number of DWDs in the corresponding structures of our Galaxy.
Following the composition of the Galaxy structure of \citet{2011MNRAS.414.2446M}, we divide our Galaxy into bulge, thin disk, and thick disk, with total masses of each component of $8.9\times10^9~M_{\sun}$, $4.32\times10^{10}~M_{\sun}$ and $1.44\times10^{10}~M_{\sun}$, respectively. In Model 1, we calculate the total number of DWDs, $N_{\rm Galaxy}$, by scaling the number ($N_{\rm fixed}$) of DWDs generated by bursts of the star formation history. In Model 2, we calculate the total number of DWDs ($N_{\rm Galaxy\_Bulge}$, $N_{\rm Galaxy\_Thindisk}$, $N_{\rm Galaxy\_Thickdisk}$) by scaling the numbers of DWDs ($N_{\rm fixed\_Bulge}$, $N_{\rm fixed\_Thindisk}$, $N_{\rm fixed\_Thickdisk}$) generated by three star formation histories for the bulge, thin disk and thick disk in Model 2. 
For the entire sample, we scale up the fixed subsets by conducting random sampling with replacement (allowing sampling of the same source more than once) from the fixed subsets.

\subsection{Galactic structure}
The position of the DWD in our Galaxy is essential for calculating the flux of the light curve (Sec.~\ref{subsec:LightCurve}) and the signal-to-noise ratio (SNR) of the GW signals (Sec.~\ref{subsec:GWsig}). Here, we assume that the structure of our Galaxy is axisymmetric, and assign a position to each pair of DWDs by sampling the distribution of the mass densities in different Galactic structures. Previous work usually assumed an exponential squared bulge and a squared hyperbolic secant plus an exponential disk or a double exponential disk~\citep{2004MNRAS.349..181N,2010A&A...521A..85Y,2019MNRAS.483.5518K}.  
A comprehensive summary of the Galactic models widely employed in the literature is provided in~\citet{2012ApJ...758..131N}. 
In this work, we adopt the density profile and the values of the associated parameters for the bulge, thin disk, and thick disk described in \cite{2011MNRAS.414.2446M}, except for the Sun-Galactic center distance, for which we adopt $R_0=8.178\,\mathrm{kpc}$ \citep{2019A&A...625L..10G}.

\subsection{Simulated DWDs}\label{sec:simDWD}

As mentioned above, we generate $N_{\rm fixed}$ DWDs as a fixed subset using different star formation histories for Model 1 and Model 2. 
To ensure adequate sampling of the short period end of DWDs, with periods of approximately several minutes, a relatively large $N_{\rm fixed}$ is necessary. After several trial experiments, we decided to set $N_{\rm fixed} = 1.7\times10^8$, so that the values of $N_{\rm Galaxy}/N_{\rm fixed}$ are 0.87, 1.72, and 5.17 for the bulge, thin disk, and thick disk, respectively. 
We then scaled up (scaled down for the bulge) the simulated fixed population of DWDs to represent the entire population expected in our Galaxy.

In addition to the intrinsic physical parameters discussed in Sec.~\ref{sec:maths}, we need to assign values for the extrinsic geometric parameters to each pair of DWDs. Specifically, the radial distance $R$ and the height $z$ in the cylindrical polar coordinates are drawn according to the density profiles for the bulge and disk, respectively. The azimuth angle $\phi$ is drawn from the uniform distribution between $[0,2\pi]$. We transform $(R,z,\phi)$ to the Ecliptic, Galactocentric, ICRS coordinates to get ecliptic longitude, ecliptic latitude, distance (lon, lat, $d$), Galactic longitude, Galactic latitude, distance($l$, $b$, $d$), Right ascension, Declination, distance (RA, Dec, $d$) for the convenience of the next steps. Here $d$ represents the distance between the DWD and the observer.

\begin{figure}
	\includegraphics[width=\columnwidth]{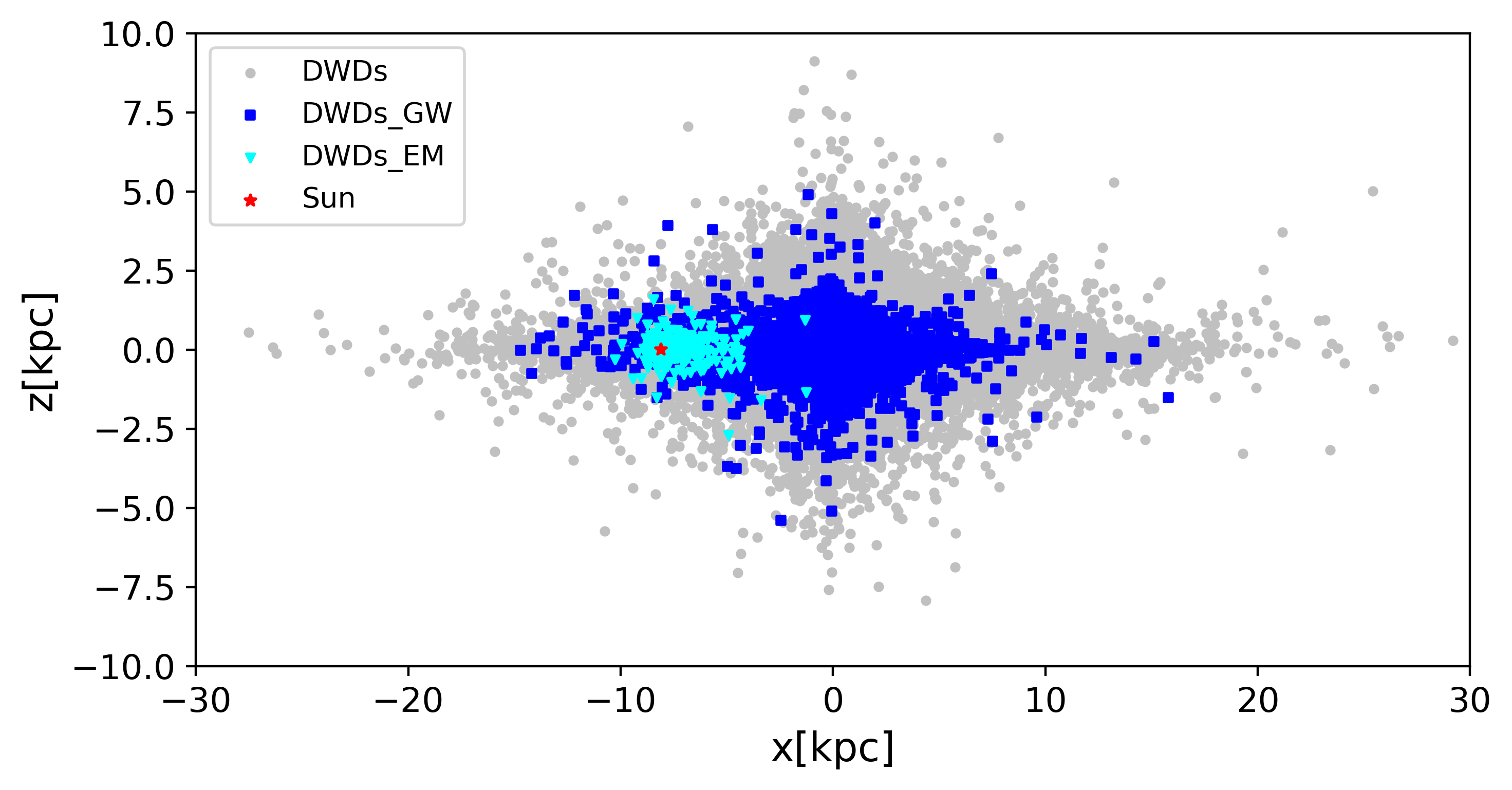}
    \caption{Distribution of DWDs in the $x-z$ plane of the Galactic Cartesian coordinate system. Grey squares represent a small representative subset of the simulated Galactic DWDs. Blue squares represent DWDs detectable by TianQin ($\mathrm{SNR} > 7$), and cyan squares represent DWDs detectable by Gaia and VRO.}
    \label{fig:Galactic_x_z}
\end{figure}

The cosine of the orbital inclination $\iota$ follows a uniform distribution between $[0,1]$. Since the detection frequency of DWDs relevant to space-borne GW detectors is mainly in the millihertz range, we select DWDs with orbital periods of less than 30 hours for further analysis. 
Note that this period is much longer than those detectable by space-borne GW detectors and can potentially be reduced to a lower value without changing the following results. However, as shown in Sec. ~\ref{sec:joint}, a fraction of the DWDs detectable by Gaia and VRO will have periods on the order of hours. 
In the end, we obtained final populations of Galactic DWDs with orbital periods less than 30 hours in Model 1 and Model 2, which amount to $6.8\times10^7$ and $4.454\times10^7$, respectively. 
In Figure \ref{fig:Galactic_x_z}, we show the distribution of a small representative subset of the simulated DWDs (grey squares) in our Galaxy, along with the ones detectable by EM and GW observations (see details in Section~\ref{sec:emobs} and \ref{sec:gwobs}.) 
\begin{table*}
	\centering
	\caption{Summary of entire DWD population in different Galactic components. He represents helium WD, CO represents carbon-oxygen WD, and ONe represents oxygen–neon–magnesium WD. ONe + WD represents the DWD in which one star is an ONe WD and the other is a WD of any type.}
	\label{table:number_DWDs}
    
        \begin{threeparttable}
        \setlength{\tabcolsep}{4.5mm}{
    	\begin{tabular*}{\linewidth}{@{}lccccccc@{}} 
                \hline
                Component& DWD type & $\mathrm{N_{total}}$ (P<30h)&$\mathrm{N_{total}}$ (TQ)&$\mathrm{N_{total}}$ (LISA)&$\mathrm{N_{total}}$ (P<30h)&$\mathrm{N_{total}}$ (TQ)&$\mathrm{N_{total}}$ (LISA)\\
  & & Model 1& (SNR>7)&(SNR>7)& Model 2&(SNR>7)&(SNR>7)\\
  \hline
   Bulge & He + He & $4.76\times10^6$& 355 & 1347&$3.37\times10^6$&345&1321\\
         & CO + He & $3.16\times10^6$& 291 & 829&$2.22\times10^6$&282&875\\
         & CO + CO & $1.20\times10^6$& 26 & 38&$9.76\times10^5$&4&13\\
         & ONe + WD & $6.64\times10^4$& 3 & 4&$6.59\times10^4$&11&23\\
   Thin disk & He + He & $2.31\times10^7$& 1676 & 6747&$7.35\times10^6$&2692&10095\\
         & CO + He & $1.54\times10^7$& 1475 & 4150&$9.74\times10^6$&4480&12946\\
         & CO + CO & $5.82\times10^6$& 104 & 207&$6.20\times10^6$&954&2838\\
         & ONe + WD & $3.21\times10^5$& 11 & 24&$5.23\times10^5$&271&633\\
   Thick disk & He + He & $7.70\times10^6$& 529 & 2211&$7.61\times10^6$&650&2561\\
         & CO + He & $5.12\times10^6$& 469 & 1350&$4.47\times10^6$&480&1471\\
         & CO + CO & $1.94\times10^6$& 25 & 46&$1.92\times10^6$&2&16\\
         & ONe + WD & $1.08\times10^5$& 5& 7&$1.28\times10^5$&9&23\\
         & total & $6.87\times10^7$& 4969 &16960&$4.45\times10^7$&10180&32815\\
    		\hline
    	\end{tabular*}}
    \end{threeparttable}   
\end{table*}

\section{EM observation}\label{sec:emobs}
 \begin{table}
    \caption{Science requirements for Gaia and VRO. The referenced parameters are from \citet{2016A&A...595A...1G} and \citet{2019ApJ...873..111I}, respectively. }
    \label{table:Gaia-LSST}
    \begin{tabular}{lll}
    \hline
  & Gaia & VRO\\
  \hline
   Survey Area & whole sky & ~20000 ${\rm deg}^2$ \\
  Single Visit Limiting Magnitude & 21 & $u$: 23.9; $g$: 25; $r$: 24.7; \\
   & &$i$: 24; $z$: 23.3: $y$: 22.1\\
  Total visits per sky patch & 70 & $u$: 56; $g$: 80; $r$: 184; \\
  & & $i$: 184; $z$: 160: $y$: 160\\
  Wavelength coverage & 330–1050 nm & 320–1050~nm~($ugrizy$)\\
  Nominal mission lifetime& 5 yr &10 yr\\
    \hline  
    \end{tabular}
\end{table}

In this section, we utilize PHOEBE~\citep[the PHysics Of Eclipsing BinariEs;][]{10.1088/978-0-7503-1287-5,2005ApJ...628..426P,2016ApJS..227...29P,2018ApJS..237...26H,2020ApJS..247...63J,2020ApJS..250...34C} to generate light curves for the simulated DWDs in detached phase in Section~\ref{sec:pop} and investigate the detection capabilities of DWDs based on the photometric performance of Gaia and VRO. The parameters of Gaia and VRO used in this paper are summarized in Table~\ref{table:Gaia-LSST}.

The Gaia mission was launched in December 2013 and has since been in operation for over 10 years~\citep{2016A&A...595A...1G}.
To obtain the most comprehensive information about our Galaxy, Gaia scans the entire sky and measures astrometry, photometry, radial velocities, and many astrophysical characteristics for billions of stars. Gaia has released several datasets containing approximately 1.8 billion sources brighter than 21 so far ~\citep{2016A&A...595A...2G,2018A&A...616A...1G,2021A&A...649A...1G,2022arXiv220800211G}. 
These progressive data releases have seen impressive growth in the identification of white dwarf candidates within the catalog, increasing from 73,221 to 359,000 \citep{2018MNRAS.480.4505J,2019MNRAS.482.4570G,2021MNRAS.508.3877G}. 
So far, most of Gaia's discoveries have been made by high precision measurements of proper motion, parallax, color, etc., which complement the discoveries made by light curves from SDSS, ASAS-SN, ATLAS, ZTF, GOTO, etc. Recently, part of the Gaia light curves have been released in Gaia DR3 along with its first catalogue of eclipsing binary candidates~\citep{2023A&A...674A..16M}.

\begin{figure}
	\includegraphics[width=\columnwidth]{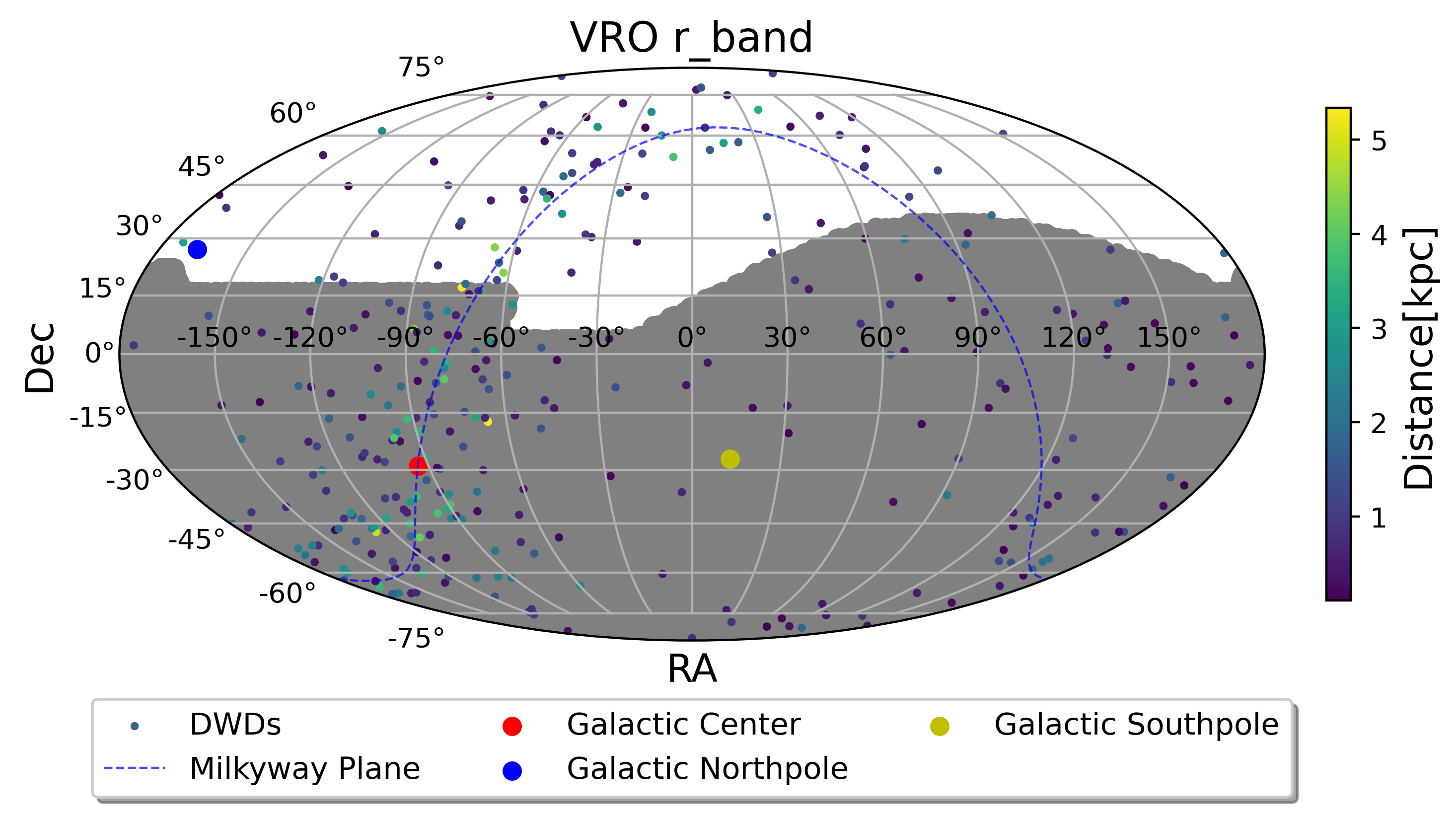}
    \caption{Distribution of detectable eclipsing DWDs in the VRO $r$-band across the celestial sphere from our simulation. The sky is shown in an equal-area Mollweide projection in equatorial coordinates (R.A. is increasing from left to right). The gray area indicates the sky region covered by VRO $r$-band.}
    \label{fig:LSST-r}
\end{figure}

The Vera C. Rubin Observatory (VRO), formally known as Large Synoptic Survey Telescope (LSST), is expected to start in 2025. VRO will produce an abundance of images and data products that will contribute to exploring our solar system, galaxy, and universe over the nominal 10 year survey plan. The telescope will observe the southern celestial hemisphere (about 20000 square degrees) in six broad-band $(ugrizy)$ filters covering the wavelength range of 320 to 1050~nm. 
As the principal component of VRO, the main wide-fast-deep (WFD) survey will pay about 825 visits to each 9.6~square degree field summed over all six bands~\citep{2017arXiv170804058L,Bianco_2022,2023ApJS..266...22S}. The number of visits per sky patch, listed in Table~\ref{table:Gaia-LSST}, serves as the baseline for our simulation. The actual surveys, however, may differ to some extent, particularly with the inclusion of deep drilling fields. 
For this study, we have employed simulated 10-year survey pointing databases generated by the Operation Simulator~\citep[OpSim;][]{2019AJ....157..151N,2016SPIE.9910E..13D,2014SPIE.9150E..15D} on NOIRLab’s Astro Data Lab\footnote[1]{\url{https://datalab.noirlab.edu/}} to get the regions of the sky covered by VRO. The shaded area in Figure \ref{fig:LSST-r} represents the sky coverage of the VRO-$r$ band. %

\subsection{Light curve simulation}\label{subsec:LightCurve}

\begin{figure}
    \centering
    \subfigure[]{
    \label{fig.lc_sub.1}
    \includegraphics[width=\columnwidth]{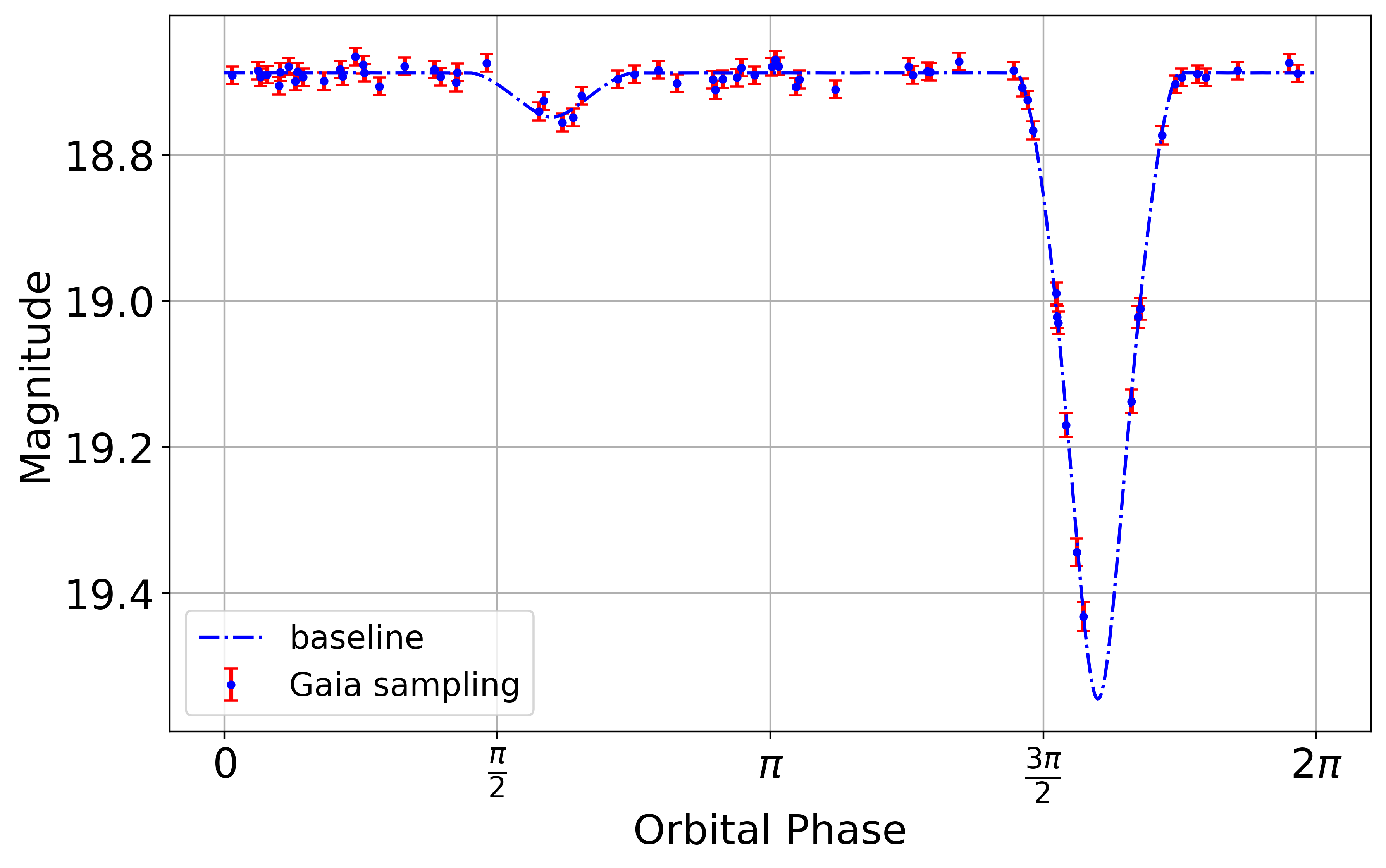}
    }\\[5pt]
    
    \subfigure[]{
    \label{fig.lc_sub.2}
    \includegraphics[width=\columnwidth]{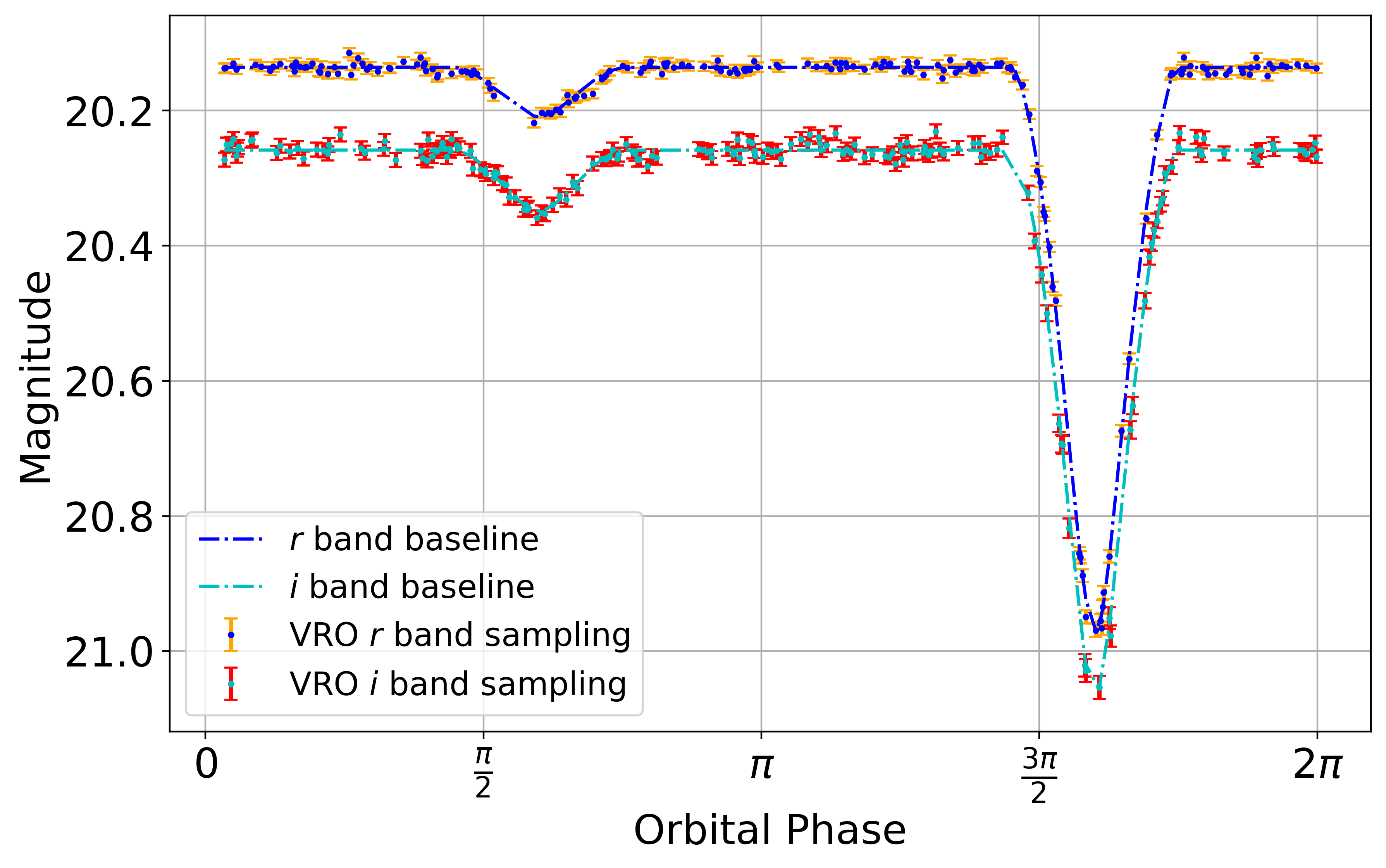}
    }\\[5pt]
    
    \caption{Simulated light curves of an eclipsing DWDs in the Gaia $G$-band (top panel), the VRO-$r$ and -$i$ bands (bottom panel). 
    Dash-dotted lines represent the simulated noiseless light curves. The number of simulated Gaia and VRO photometric data points are taken from Table~\ref{table:Gaia-LSST}. Note that the light curves shown here are phase folded. See the last paragraph of this subsection for the parameters used in this example.
    }
    \label{fig:lc}
\end{figure}

PHOEBE can produce and fit light curves and radial velocity profiles of eclipsing binaries. 
In light curve simulation, the input parameters for the binary include: $r_1$, $r_2$, $\mathrm{T_1}$, $\mathrm{T_2}$, $a$, $d$, $\mathrm{P}$, $\iota$, and $e$, where $r_1$ and $r_2$ are the radii and $\mathrm{T_1}$ and $\mathrm{T_2}$ are the effective temperatures of each white dwarf, $a$ is the orbital semi-major axis, $d$ represents the distance of the DWD from the observer, $\mathrm{P}$ is the orbital period, $\iota$ is the orbital inclination, and $e$ is the orbital eccentricity. In addition to the primary influence of these intrinsic parameters on the light curve, we also consider secondary effects, such as limb-darkening and mutual irradiation effects. 
The limb-darkening coefficient is automatically obtained by interpolating the applicable tables based on the atmospheric parameters for ease of calculation \citep{2016ApJS..227...29P}. For the mutual irradiation effects, we adopt the default irradiation method `horvat'~\citep{2019ApJS..240...36H}. 
However, we have not considered the surface distortion due to tides and rotation (ellipsoidal variation) in this work. 
For the system shown in Fig.~\ref{fig:lc}, the differences in the mean magnitudes between the case with and without the distortion are less than 0.01 mag for Gaia and VRO. Thus, ignoring the distortion would be appropriate for estimating the number of detectable DWDs. In contrast, a more accurate modeling of the light curve would be needed to estimate the parameters, especially for the short period system.
Finally, we convert the output flux measured in $\mathrm{W/m^2}$ to apparent magnitudes for various bands in Gaia, VRO, and include the Gaussian noise from observational uncertainties and the extinction effect due to interstellar dust in our synthetic light curves. 

For Gaia, we calculate the magnitude according to the definition of \cite{2018MNRAS.479L.102C} and use the photometric standard error of a single visit: 
\begin{equation}
\sigma_G =1.2\times10^{-3}\left(0.04895 z_{x}^2+1.8633 z_{x}+0.0001985\right)^{1/2} \,, \label{eq:5}
\end{equation}
where $z_{x}=\max \left[10^{0.4(12-15)}, 10^{0.4(G-15)}\right]$~\citep{2016A&A...595A...1G}. 

Similarly, we calculate the VRO magnitude in the AB magnitude system as used in \cite{2009arXiv0912.0201L}. The photometric error for a point source in a single visit can be obtained using the following relation:

\begin{equation}
\sigma_L^2=\sigma_{\rm sys}^2+\sigma_{\text {rand }}^2  \,,\label{eq:7}
\end{equation}
where the systematic photometric error $\sigma_\mathrm{sys}=0.005\ \mathrm{mag}$ and the random photometric error $\sigma_{\rm rand}^2=(0.04-\gamma) x+\gamma x^2\left(\mathrm{mag}^2\right)$ with $x=10^{0.4(m-m_5)}$. $m_5$ is the $5\sigma$ depth (for point sources) in a given band. The values of the parameters can be found in Table 3.2 of \cite{2009arXiv0912.0201L}.

We adopt the distance-dependent extinction model defined in \citet{2018ApJ...854L...1B} for the DWDs distributed in the Galactic disk and bulge. We first calculate the $V$-band extinction and then convert it to the Gaia $G$-band and the VRO $(ugrizy)$ bands using the relationship developed in~\citet{1989ApJ...345..245C}. 

Using the above prescription, we simulate the light curves of eclipsing DWDs observable by Gaia and VRO. 
In Figure~\ref{fig:lc}, we show the simulated light curves with phase folded for an eclipsing DWD using the following parameters: $r_1= 0.02980\ \mathrm{R}_{\sun}$, $r_2= 0.0275\ \mathrm{R}_{\sun}$, $\mathrm{T_1}= 26300\ \mathrm{K}$, $\mathrm{T_2}= 9200\ \mathrm{K}$, $a= 0.1227\ \mathrm{R}_{\sun}$, $d= 1.6\ \mathrm{kpc}$, $\mathrm{P}= 0.1466\ \mathrm{h}$, and $\iota= 82^{\circ}$. The primary eclipse is clearly visible in the simulated light curves.

\subsection{Source selection}\label{sec:detected}

From all simulated DWDs, we select detectable eclipsing DWDs in the Gaia and VRO surveys using criteria that include (i) the survey limiting magnitudes, (ii) the target coordinates, (iii) the $\chi^2$ tests of the light curves, and (iv) the inclination angles of the DWDs. For Gaia and VRO, a detectable eclipsing DWD must meet all these four criteria simultaneously.

For Gaia, we select DWDs that are brighter than magnitude 21 over the entire sky. 
For VRO, we select DWDs based on the sky coverage and the limiting magnitude of each bandpass. 
The sky map of VRO is shown in Figure~\ref{fig:LSST-r}, and the limiting magnitudes for each bandpass are listed in Table~\ref{table:Gaia-LSST}. 
For Gaia and VRO, to ensure the selected binaries are eclipsing, we also apply the criterion of $\cos\iota\leq(r_1+r_2)/a$, where $\iota$, $r_i$, $a$ are the orbital inclination, radii of WDs, and orbital separation, respectively.

To distinguish eclipsing DWDs from non-variable stars with photometric fluctuations, we used the aforementioned $\chi^2$ test similar to that of~\cite{2019MNRAS.483.5518K} to select DWDs with $\chi^2>3$. Here,  
\begin{equation}\label{eq:chi2}
\chi^{2}=\frac{1}{N} \sum_{i=1}^{N}\left(\frac{M_{i}-M_{\text {mean}}}{\sigma_{i}}\right)^{2}  \,, 
\end{equation}
where $N$ is the total number of visits per star given in Table~\ref{table:Gaia-LSST}, $M_{i}$ represents the measured photometric magnitude from each visit, $\sigma_{i}$ represents the photometric error as stated in Equations~\ref{eq:5} and~\ref{eq:7} for Gaia and VRO, respectively, and $M_{\text{mean}}$ represents the mean photometric magnitude for all visits in the same bandpass. Based on the above selection criteria, the numbers of detectable eclipsing DWDs in our simulations are counted and summarized in Table \ref{table:number}. 

\begin{table}
    \caption{The number of detectable eclipsing DWDs in our Galaxy by Gaia and VRO in our simulations. For VRO, we list the number of detection for each bandpass in the VRO filters $(ugrizy)$, as well as the total number obtained when combining all bandpasses (the second last column). We also derive the total number of detections when combining observations from Gaia and VRO (the last column). } \label{table:number}
    \begin{tabular}{ccccccccccc}
    \hline
  & Gaia & \multicolumn{7}{|c|}{VRO}&total\\
  & & $u$ &$g$ &$r$ &$i$ &$z$ &$y$ &total \\
  \hline
    Model 1&67&16&189&247&116&55&1&273&289\\
    \hline
    Model 2&186&38&429&488&228&107&3&554&623\\
    \hline
    \end{tabular}
\end{table}

\begin{figure}
    \centering
    \includegraphics[width=\columnwidth]{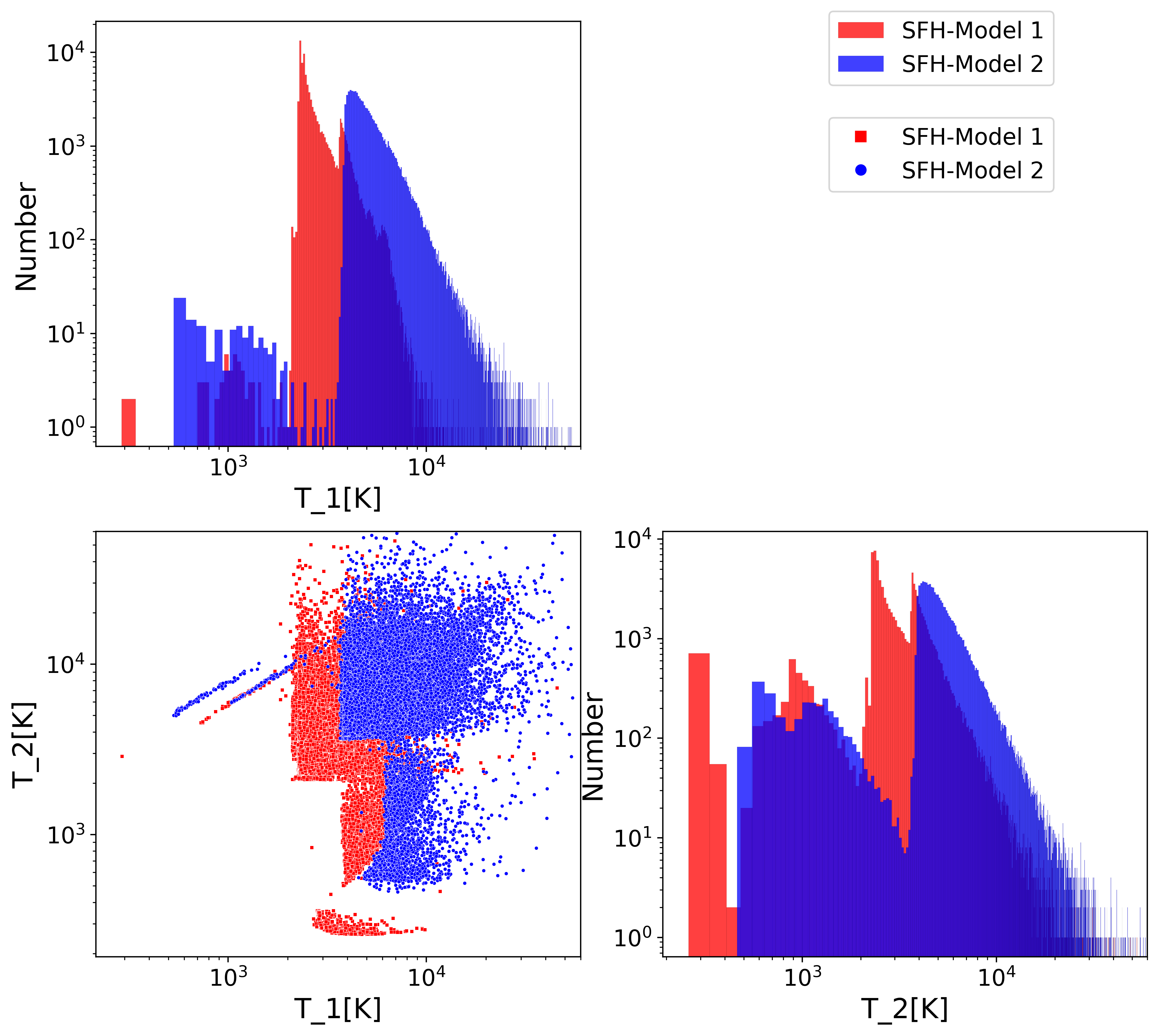}
    \caption{Corner plots for the temperatures of the primary ($T_1$) and companion ($T_2$) stars of DWDs from simulations in Model 1 (red) and Model 2 (blue) of star formation history.}
    \label{fig:temperature}
\end{figure}

\begin{figure}
    \centering
    \includegraphics[width=\columnwidth]{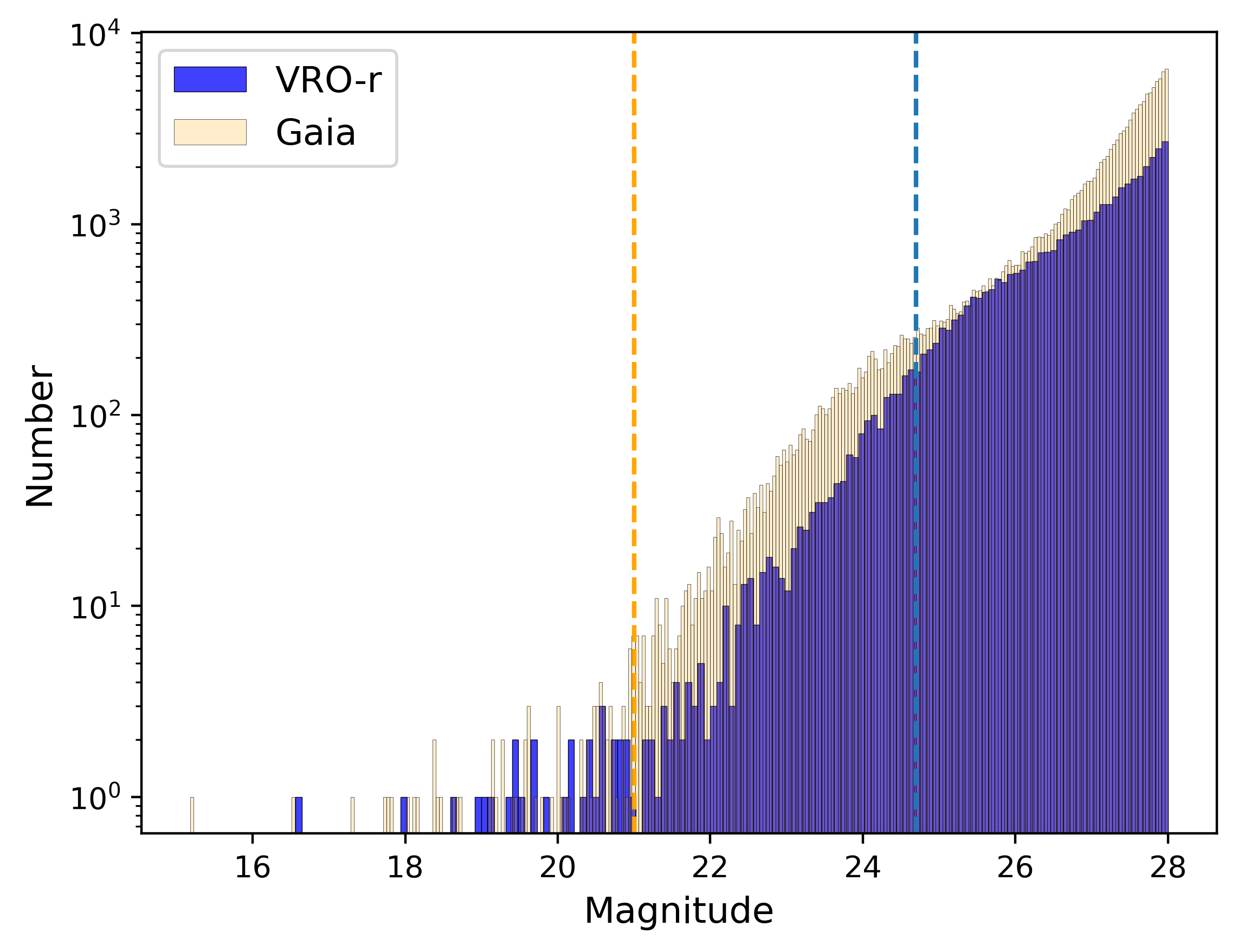}
    \caption{Distributions of the mean apparent magnitudes of the eclipsing DWDs in Model 1. The light orange and blue histograms represent samples from the Gaia $G$-band and the VRO-$r$ band, respectively. The orange and blue vertical dash lines represent the limiting magnitudes of the Gaia $G$-band and the VRO-$r$ band, respectively.}
    \label{fig:Mag_dis}
\end{figure}

\begin{figure*}
    \centering
    \includegraphics[width=2\columnwidth]{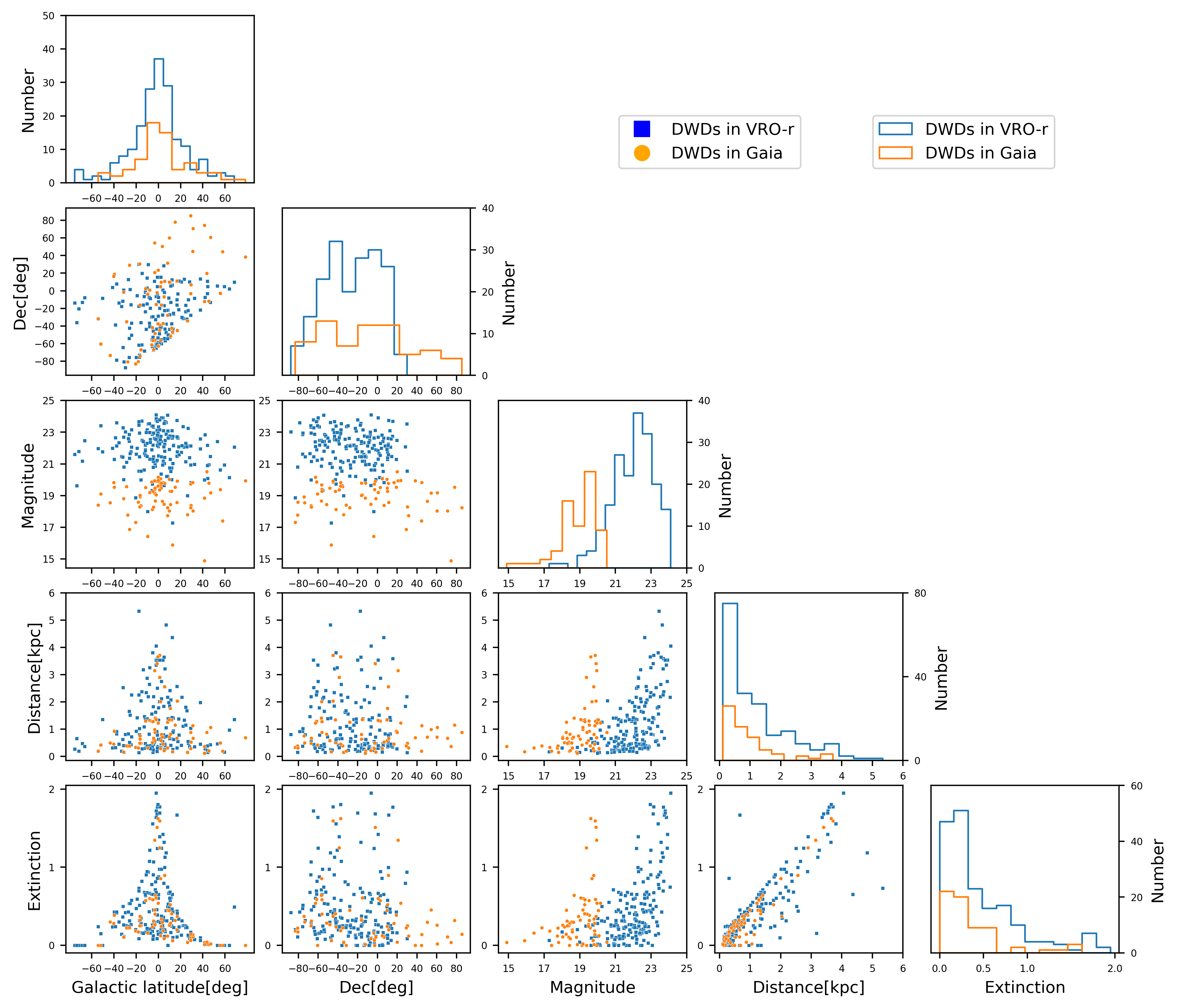}
    \caption{The corner plot for several parameters of eclipsing DWDs in the Gaia and VRO-$r$ band samples. The diagonal histograms show the distributions of the Galactic latitude, Declination (Dec), apparent magnitude, distance, and extinction. The scatter plots show the correlations between each pair of parameters.
}
    \label{fig:pairplot_GL}
\end{figure*}

\begin{figure*}
    \centering
    \subfigure[Gaia]{
    \label{fig.Gaia-LSST-sky_sub.1}
    \includegraphics[width=\columnwidth]{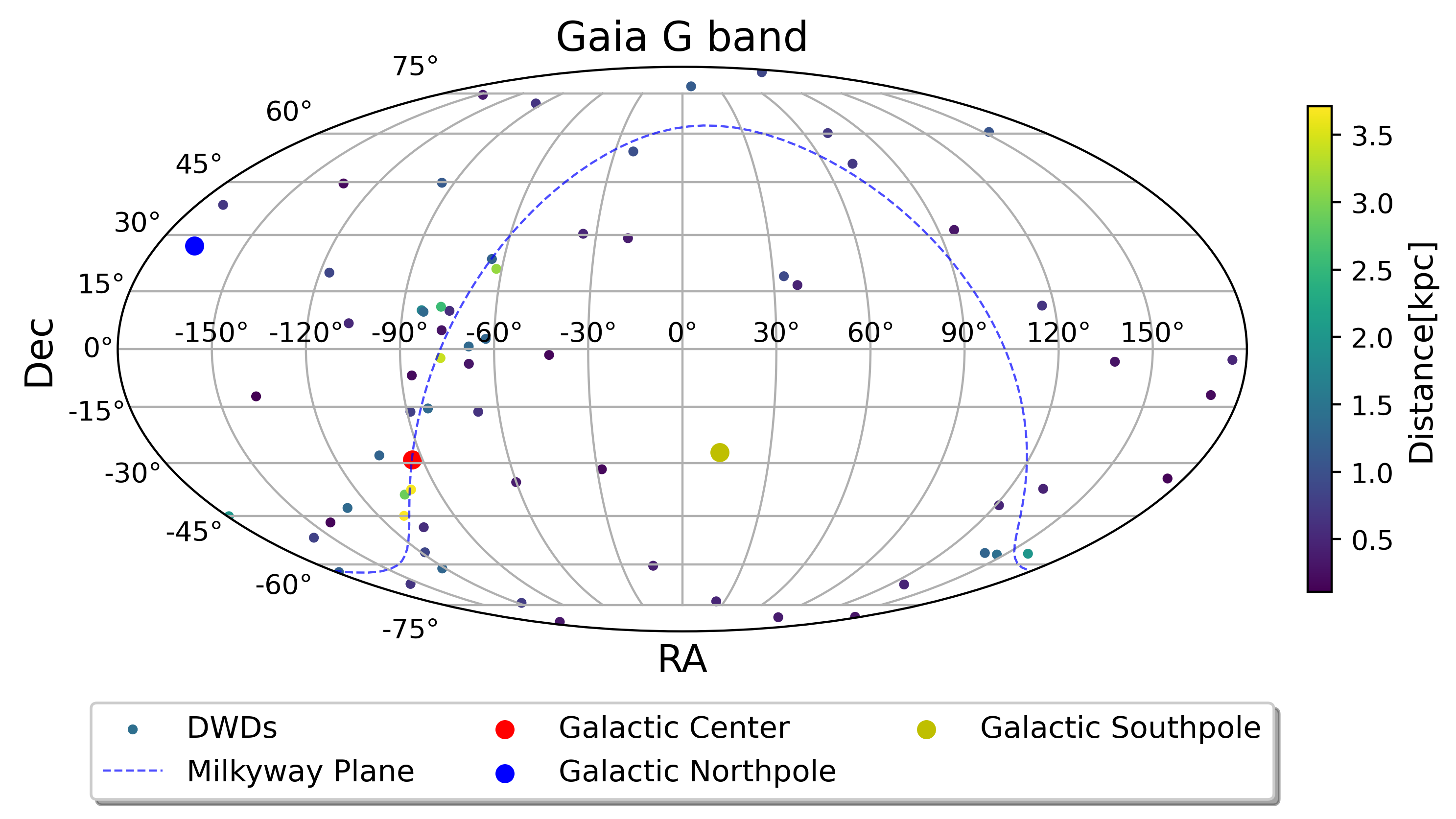}
    } 
    \subfigure[VRO]{
    \label{fig.Gaia-LSST-sky_sub.2}
    \includegraphics[width=\columnwidth]{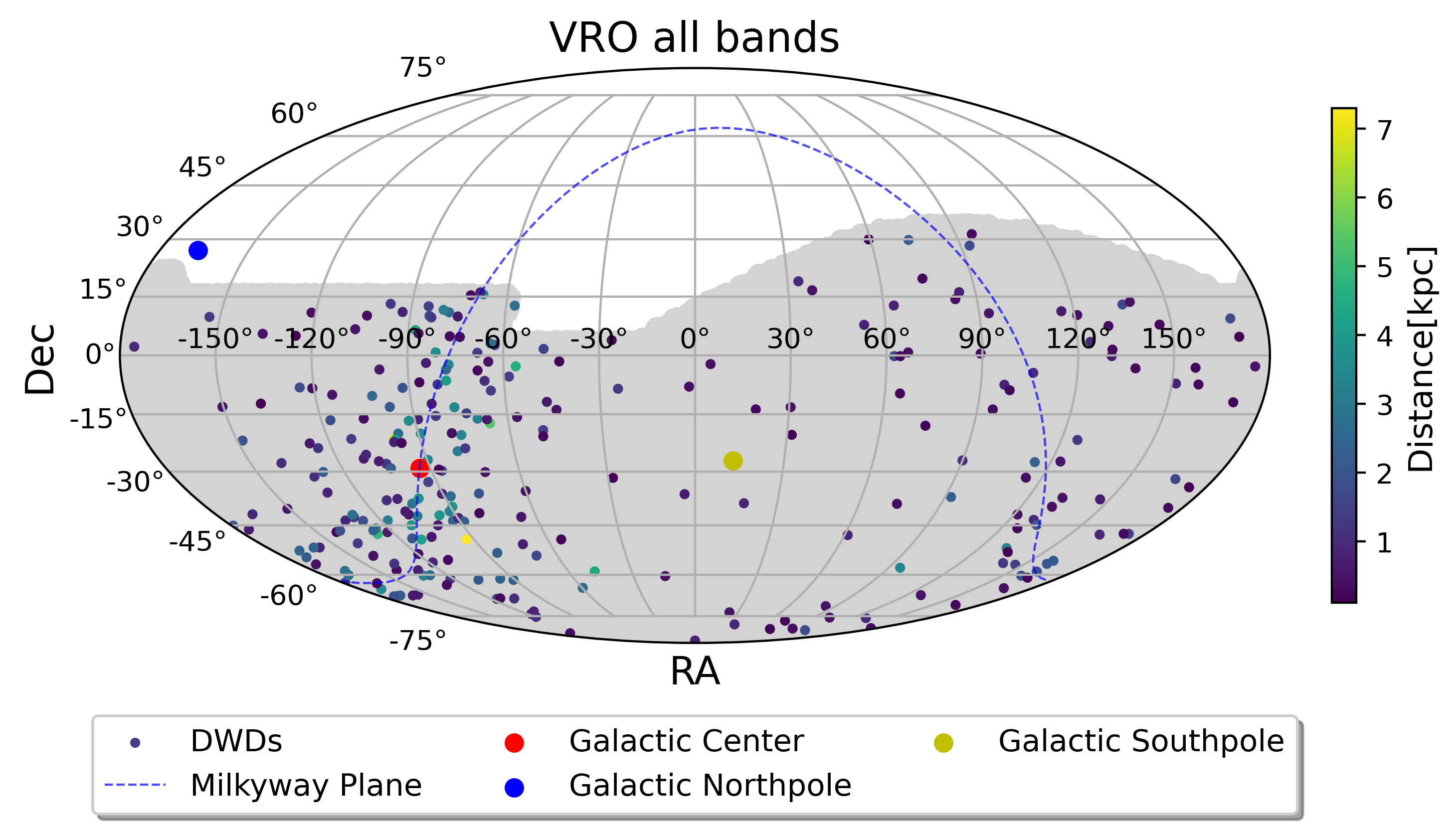}
    }
    \caption{Sky distributions of the eclipsing DWDs (Model 1) detectable by Gaia (left panel) and VRO (right panel).}
    \label{fig:Gaia-LSST-sky}
\end{figure*}

\subsection{Results}

In Table~\ref{table:number}, we show the number of eclipsing DWDs detectable by Gaia and VRO in both Model~1 and Model~2. Our results show that 67 and 273 DWDs produced in Model~1 can be detected by Gaia and VRO, respectively. In Model 2, these numbers increase to 186 and 554, respectively. 
The primary factor contributing to the difference between Model 1 and Model 2 is the choice of the star formation history as described in Section~\ref{sec:pop}, which results in different temperature distributions for the simulated populations of DWDs as shown in Figure~\ref{fig:temperature}. 
Due to the longer period of cooling timescale, the DWDs from Model 1 have overall lower temperature distributions for the primary and companion stars and are therefore less likely to be observed compared to those from Model 2.
The distributions of mean apparent magnitudes for the corresponding DWDs are shown in Figure \ref{fig:Mag_dis}. 
The mean apparent magnitude refers to the sum of apparent magnitudes at each sampling point on the light curve divided by the number of sampling points (visits). The magnitude at each sampling point takes into account photometric error and extinction. Therefore, although many sources have a mean magnitude below the limiting magnitude, some of their sampling points (especially those during eclipse) exceed the limiting magnitude, and are thus excluded from our analysis. Furthermore, the photometric error bar increases with increasing magnitude, so more higher magnitude sources have a greater chance of being excluded in VRO than in Gaia.
As we can see, only a small fraction of the DWDs fall within the limiting magnitude, which is consistent with the results reported by~\citet{2009CQGra..26i4030N}. 

Figure~\ref{fig:pairplot_GL} shows the distributions of the main parameters of the DWDs detectable by the Gaia and the VRO-$r$ band for Model 1. According to the distribution of Galactic latitude, about $70\%$ of the DWDs are located within the $\pm 20$ degree range of the Galactic plane. In the equatorial coordinate system, compared to the Gaia's all-sky observations, the DWDs from the VRO observations are mainly located in the southern sky region. From the magnitude distribution, we can see that the magnitude increases with increasing distance. The peak of the magnitude distribution of the DWDs observed by VRO is around 22, which is slightly higher than that of Gaia. Note that from the distribution shown in the top panel of the third column, the VRO extends to fainter DWDs compared to Gaia. 
This is mainly because the Gaia $G$-band covers a wider wavelength range of $330~\rm{nm}-1050~\rm{nm}$, while the VRO-$r$ band covers a range of $560~\rm{nm}-706~\rm{nm}$. As a result, the Gaia $G$-band collects more photons, resulting in a higher integrated flux. In this work, we used PHOEBE to simulate the flux and converted it to magnitudes. As expected, the flux simulated by PHOEBE in the Gaia $G$-band is higher than in the VRO-$r$ band for the same source. This can already be seen in Fig.~\ref{fig:lc}, where the same source can have a clear difference in magnitude between the Gaia $G$-band and VRO-$r$ band. Therefore, the bright sources in Gaia will appear less bright in VRO, resulting in fewer bright DWDs in the VRO-$r$ band.
The distribution of distances shows that the DWDs detectable by VRO are mainly within 4 kpc and those detectable by Gaia are mainly within 2 kpc. 
As we can see from  Figure~\ref{fig:Galactic_x_z} and Section~\ref{sec:gwobs}, the space-borne GW detectors will have a greater distance reach than these optical telescopes, and therefore can detect more DWDs in our Galaxy.
For the EM observations, extinction is a critical factor in the calculation of the apparent magnitude, which is mainly determined by the Galactic latitude, distance, and structure of the Galaxy. The closer the DWDs are to the Galactic plane, the stronger the extinctions. The distance of the DWDs from the observer and the extinction show a proportional linear relationship; the greater the distance from the DWDs, the stronger the extinction.

Figure~\ref{fig:Gaia-LSST-sky} illustrates the sky distributions of 67 (273) eclipsing DWDs in Model 1 that are detectable by Gaia and VRO.
Our results show that VRO can detect 273 eclipsing DWDs in Model 1 and 554 in Model 2. 
These numbers are lower than those reported by \citet{10.1093/mnras/stx1285} (hereafter referred to as K17), which are 1100 and 1475 for two different models. 
Several factors account for this discrepancy. 
The major factor is the calculation of errors of apparent magnitude of the WDs. K17 effectively uses $\sigma_{\rm{rand}}^4$ instead of $\sigma_{\rm{rand}}^2$ in Eq.~\ref{eq:7}, which leads to an underestimation of the photometric errors, especially for fainter WDs.
In our validation tests, using $\sigma_{\rm{rand}}^4$ for the random
photometric errors allows VRO to detect 998 DWDs in Model 1 and 1815 in Model 2, which are more consistent with the numbers reported in K17. 
In addition, our simulation of the light curves for each VRO bandpass $(ugrizy)$ is based on the number of sampling points (visits) listed in Table~\ref{table:Gaia-LSST}. In contrast, K17 considers only the VRO-$r$ band and assigns 1000 sampling points to it. 
The other minor factors include different star formation histories, initial mass functions, and the masses and mass density distributions of the Galactic disk and bulge adopted in our study.

\section{GW observation}\label{sec:gwobs}

In this section, we simulate the GW signals from the DWDs obtained in Section~\ref{sec:simDWD} and calculate their SNRs for TianQin, LISA, and their joint observation. In addition, we evaluate estimation accuracy for the parameters of the DWDs using the Fisher information matrix (FIM). 
                                            
\subsection{GW signals}\label{subsec:GWsig}

The GW strain signal measured by a detector can be expressed as:
\begin{equation}
h(t)= F^{+}(t) h_{+}(t) + F^{\times}(t) h_{\times}(t) \,.
\end{equation}
Here, $h_{+}(t)$ and $h_{\times}(t)$ are the two polarizations of the GWs emitted by a DWD, which can be calculated using the quadrupole-moment formula~\citep{1963PhRv..131..435P}. 
$F^{+}(t)$ and $F^{\times}(t)$ are the detector's antenna pattern functions which depend on the source's location and its orientation relative to the detector.  We adopt the expressions of the detector's antenna pattern functions for TianQin from~\citet[][]{2018CQGra..35i5008H,2020PhRvD.102f3021H},  and for LISA from \citet{2003PhRvD..67j3001C}. The squared SNR of the signal can be calculated as follows~\citep{1992PhRvD..46.5236F} 
\begin{equation}
\rho^2 = \frac{2}{\tilde{S}_n\left(f_0\right)} \int_0^T \mathrm{~d} t h(t)^2 \,,
\end{equation}
where $\tilde{S}_n\left(f_0\right)$ is the single-sided power spectral density (PSD) of detector at GW frequency $f_0$, and $T$ is the effective observation time. The PSD for TianQin and LISA can be found in~\citet{2018CQGra..35i5008H,2020PhRvD.102f3021H} and \citet{Robson_2019}, respectively.

The accumulation of SNR depends on the effective observation time $T$. For TianQin, we adopt an observation scheme of three months on and three months off over a nominal five-year mission duration, which corresponds to an effective observation time of 2.5 years \citep{TianQin2016,2020PhRvD.102f3021H}.
For LISA, the current baseline employs a duty cycle of 0.75 over a nominal four-year mission duration, resulting in an effective observation time of 3 years \citep{2022GReGr..54....3A}. 
Note that for combined observation, the SNRs for TianQin and LISA are added quadratically, since the noise are independent for the two detectors~\citep{2016PhRvD..94h1101T}.

\subsection{Fisher information matrix}

For multi-messenger astronomy, particularly for the EM follow-up of the GW signals, it is crucial to forecast the estimation accuracy for the parameters of the detectable DWDs. To achieve this, we use the Fisher information matrix defined as~\citep{1994PhRvD..49.2658C} 
\begin{equation}\label{eq:Fisher}
\Gamma_{i j}=\left(\frac{\partial h}{\partial \xi_i} \mid \frac{\partial h}{\partial \xi_j}\right) \,,
\end{equation}
where $\left(|\right)$ denotes the noise-weighted inner product. $\xi_i$ represents the $i$-th unknown parameter in the parameter set that includes $\mathcal{A}, \iota, \psi, \theta, \phi, f, \phi_{0}, \dot{f}$. The inverse of FIM is equal to the covariance matrix, $\Sigma_{ij}=\Gamma^{-1}_{ij}$. The diagonal element $\Sigma_{ii}$ represents the variance of the estimation error for $\xi_i$, and the non-diagonal element $\Sigma_{ij}$ represents the covariance of the estimation errors between $\xi_i$ and $\xi_j$. Based on the error propagation equation, we estimate the error of the source sky position measurement, the relative error of the chirp mass, and the relative error of the distance with the following equations:
\begin{equation}
    \Delta \Omega =2 \pi\left(\Sigma_{\theta \theta} \Sigma_{\phi \phi}-\Sigma_{\theta \phi}^2\right)^{1/2} \,,
\end{equation}
\begin{equation}\label{eq:deltdMc}
    \frac{\Delta{\mathcal{M}}}{\mathcal{M}} =\sqrt{\left(\frac{11}{5} \frac{\Sigma_{ff}}{f}\right)^2+\left(\frac{3}{5} \frac{\Sigma_{\dot{f}\dot{f}}}{\dot{f}}\right)^2+\frac{33}{25}\left(\frac{\Sigma_{ff}}{f}\right)\left(\frac{\Sigma_{\dot{f}\dot{f}}}{\dot{f}}\right) \Sigma_{f\dot{f}}} \,, 
\end{equation}
\begin{equation}\label{eq:deltdD}
    \frac{\Delta{\mathcal{D}}}{\mathcal{D}} =\sqrt{\left(\frac{\Sigma_{ff}}{f}\right)^2+\left(3 \frac{ \Sigma_{\dot{f}\dot{f}}}{\dot{f}}\right)^2+\left(\frac{\Sigma_{\mathcal{A}\mathcal{A}}}{\mathcal{A}}\right)^2} \,. 
\end{equation}
Note that DWDs are typically quasi-monochromatic sources in the mHz band, so not all DWDs in our simulation exhibit a measurable $\dot{f}$ (or $\Sigma_{\dot{f}\dot{f}}$). Here we adopt the commonly used criterion $\dot{f}\,T^2 \rho > 1$ \citep{2012PhRvD..86l4032A} to select sources for which we can obtain reliable results from Eqs.~\ref{eq:deltdMc} and \ref{eq:deltdD}. If this criterion is not met, $\dot{f}$ will be excluded from in the parameters to be estimated in our analysis. 

\subsection{Results}

Table \ref{table:number_GW} lists the expected numbers of DWDs in Model 1 and Model 2 with $\rho \geq 7(5)$ for TianQin, LISA, and their joint observation. 
For TianQin, we consider nominal mission duration of 1, 2, and 5 years, which correspond to effective observation times of 0.5, 1, and 2.5 years, respectively. For LISA, we consider nominal mission duration of 1, 2, and 4 years, which correspond to effective observation times of 0.75, 1.5, and 3 years, respectively. For Model 1, TianQin can detect approximately $5\times10^3$ DWDs with SNRs greater than 7 in 2.5 years of effective observation time and LISA can detect approximately $1.7\times10^4$ DWDs with SNRs greater than 7 in 3 years of effective observation time. For Model 2, TianQin can detect approximately $1\times10^4$ DWDs with SNRs greater than 7 in 2.5 years of effective observation time, which is comparable to the results in~\citet{2020PhRvD.102f3021H} based on a DWD population generated from SeBa~\citep{2017A&A...602A..16T}; LISA can detect approximately $3\times10^4$ DWDs with SNRs greater than 7 in 3 years of effective observation time, which is comparable to the results in~\citet{10.1093/mnras/stx1285}. The numbers of DWDs generated from Model 2 are about twice as many as these from Model 1. This is mainly due to Model 2 generating more DWDs with frequencies in the millihertz. In addition, Model 1 can serve as a lower bound for the models based on astrophysical star formation histories, because a fraction of the DWDs from other models would have already evolved in the burst-like star formation history in Model 1. The joint observation of TianQin and LISA can detect an additional 2,000 DWDs compared to LISA alone.  

\begin{figure}
	\includegraphics[width=\columnwidth]{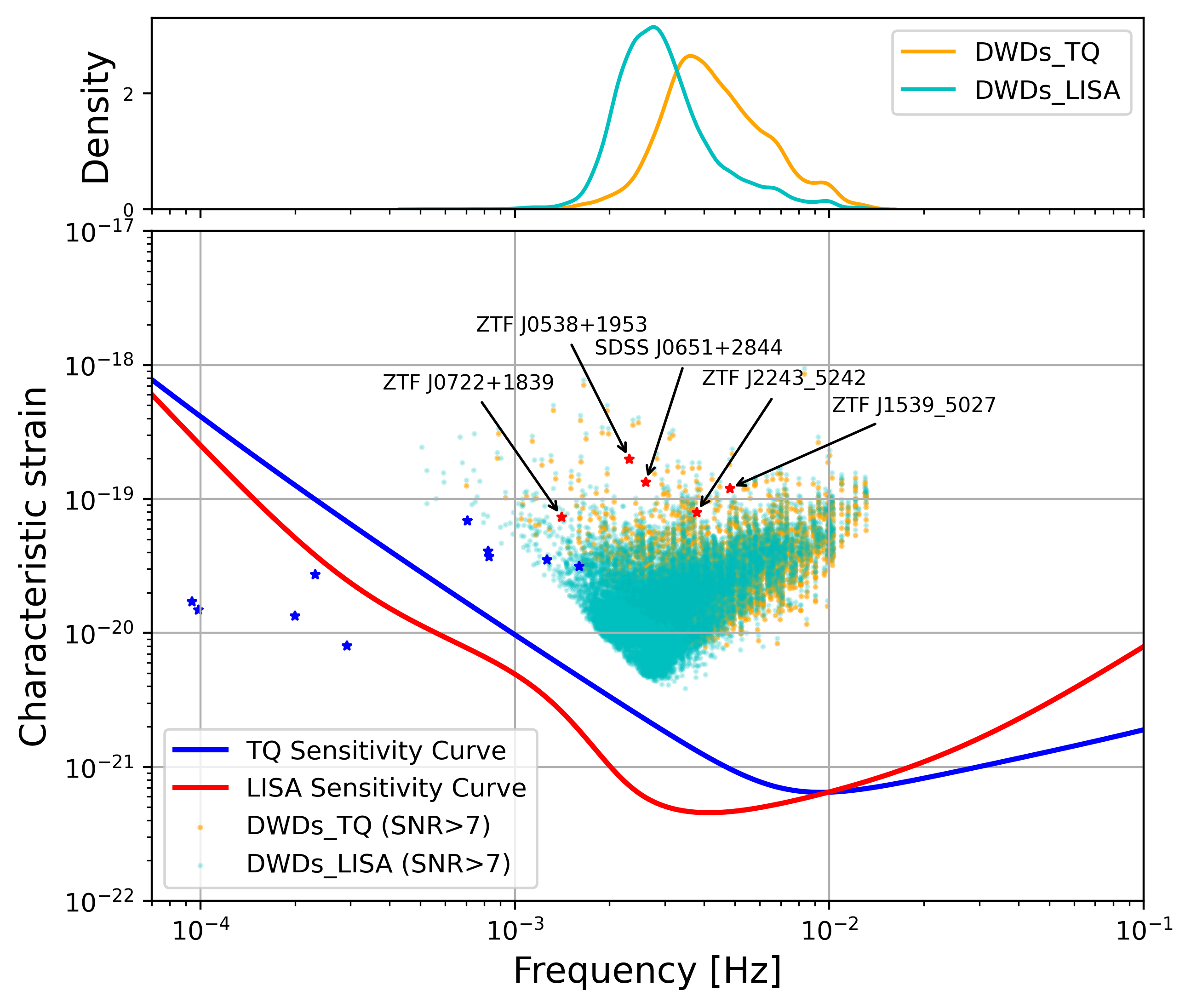}
    \caption{Bottom panel: The characteristic strain of the 15 optically observed eclipsing DWDs (indicated by blue and red stars), along with the characteristic strain of the simulated DWDs (cyan and orange dots) with $\rho >7$ for TianQin and LISA as a function of frequency. The red and blue solid lines represent the sensitivity curves of LISA and TianQin respectively. Top panel: Frequency distributions of simulated DWDs with $\rho >7$. Details on these 15 sources can be found in~\citep{2020ApJ...905L...7B,2014ApJ...780..167K,2011ApJ...735L..30P,2011ApJ...737L..16V,2011ApJ...737L..23B,Kilic:2013zma,2016MNRAS.458..845H,2017ApJ...847...10B,2020ApJ...905...32B,10.1093/mnras/stad2347,2023ApJ...950..141K}.}
    \label{fig:TQ_LISA_SNR_GE7}
\end{figure}

From the top panel of Figure \ref{fig:TQ_LISA_SNR_GE7}, we can see that the frequencies of the DWDs detectable by TianQin and LISA are predominantly in the range of $10^{-3}-10^{-2}$~Hz. The frequency distribution of DWDs detectable by LISA (centered around $2.5 \times 10^{-3}$ Hz) exhibits a slightly lower frequency peak compared to TianQin (centered around $3.5 \times 10^{-3}$ Hz). This is primarily attributed to the lower sensitivity curve of LISA compared to TianQin for frequencies below $10^{-2}$ Hz, allowing LISA to detect a greater number of lower frequency DWDs.
The bottom panel shows the sensitivity curves ($\sqrt{f{\tilde{S}_n\left(f\right)}}$) for TianQin and LISA. Among the 16 optical candidates of eclipsing DWDs (marked as stars), we found that 5 of them (marked as red stars) accumulate SNRs greater than 7 over a nominal 4-year mission duration of LISA. 

\begin{figure*}
    \centering
    \includegraphics[width=2\columnwidth]{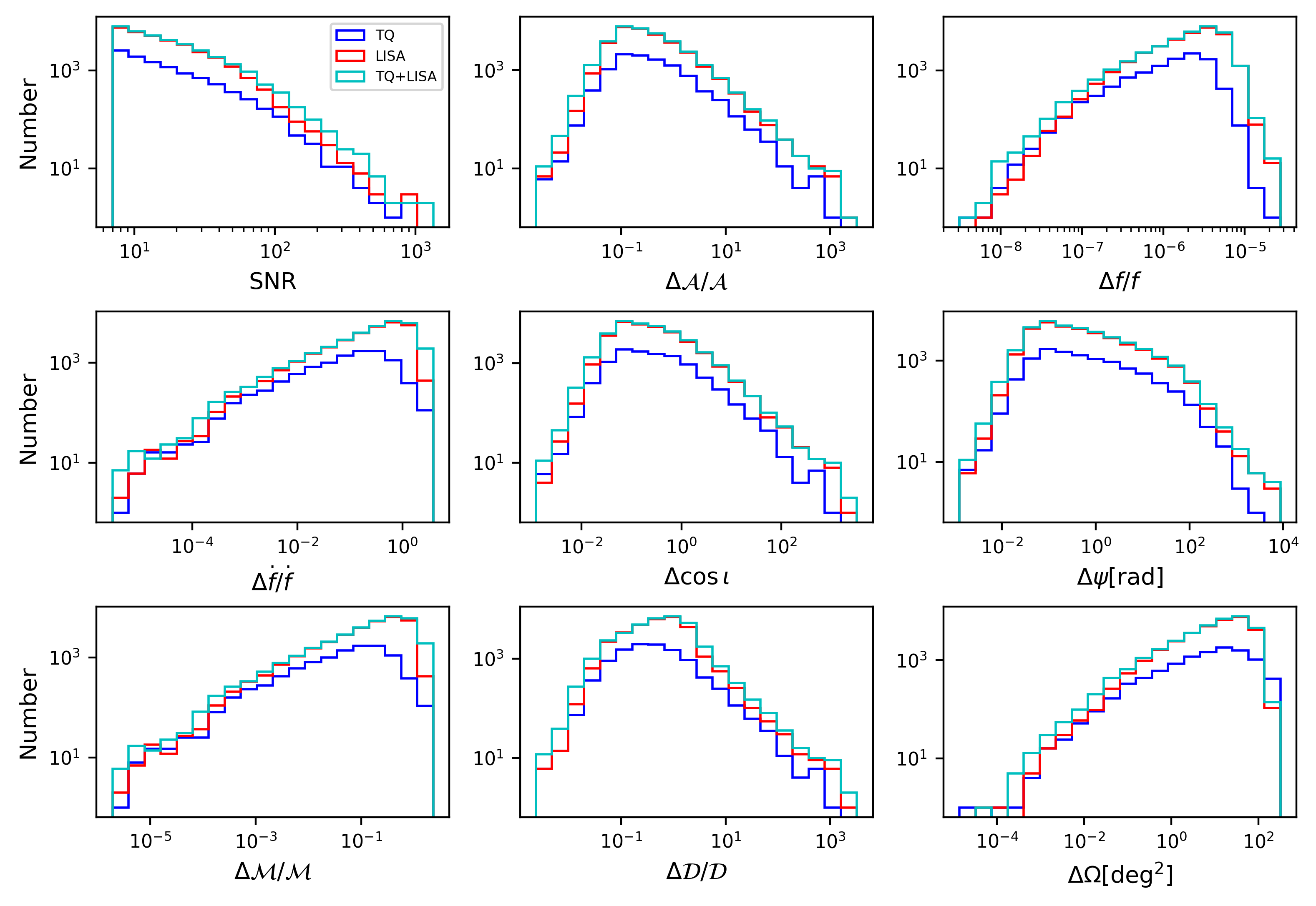}
    \caption{Histograms of the SNRs and parameter estimation accuracy of the DWDs detectable by TianQin, LISA, and their joint observation in Model 2.}
    \label{fig:FIM_TL}
\end{figure*}

Figure \ref{fig:FIM_TL} shows the distributions of parameter estimation accuracy obtained through the FIM analysis for Model 2. Our result yields 3241 (32\%), 9760 (29\%), and 10356 (29\%) DWDs with SNRs between 7 and 10 for TianQin, LISA, and their joint observation, respectively. Only 2\% of the DWDs exhibit SNRs greater than 100 for these three detector configurations. In Table \ref{tab:cum_FIM}, we count the cumulative numbers and corresponding percentages of DWDs with parameter estimation relative errors less than 0.1, 0.2, and 0.5 for TianQin, LISA, and their joint observation in Model 2. 

The relative error for distance $\Delta{\mathcal{D}}/\mathcal{D}$ is greater in comparison to the other parameters. From Equation \ref{eq:deltdD} and the distributions of $\Delta{\mathcal{A}}/\mathcal{A}$, $\Delta{f}/f$, and $\Delta{\dot{f}/\dot{f}}$ in Figure \ref{fig:FIM_TL}, we can see that the contribution to $\Delta{\mathcal{D}}/\mathcal{D}$ is predominantly attributable to $\Delta{\mathcal{A}}/\mathcal{A}$. In Model 2, approximately $79\%$ and $76\%$ of the DWDs exhibit relative errors in $\Delta{\mathcal{A}}/\mathcal{A}$ and $\Delta{\mathcal{D}}/\mathcal{D}$ exceeding $1$ for TianQin, respectively. 
For the sky localization, the numbers of the DWDs with position errors $\Delta{\Omega}<1~\mathrm{deg}^2$ ($10~\mathrm{deg}^2$) are 1889 (4974), 4010 (13750), and 4722 (14746) for TianQin, LISA, and their combination, respectively. 
Given the fields of view of Gaia and VRO are 1.26~$\rm{deg}^2$ and 9.6~$\rm{deg}^2$, respectively, these sources can be easily targeted by the EM follow-up observations.

\begin{table*}
	\centering
	\caption{Number and percentage of cumulative distributions of the relative errors of the parameters less than $0.1$, $0.2$, $0.5$ for TianQin, LISA and their joint observation in Model 2.}
	\label{tab:cum_FIM}

        \begin{threeparttable}
        \setlength{\tabcolsep}{2.5mm}{
    	\begin{tabular*}{\linewidth}{@{}cccccccccc@{}} 
    		\hline
    		&\multicolumn{3}{c}{$<0.1$}&\multicolumn{3}{c}{$<0.2$}&\multicolumn{3}{c}{$<0.5$}\\
                &TQ&LISA&TQ+LISA&TQ&LISA&TQ+LISA&TQ&LISA&TQ+LISA\\
                \hline
                $\Delta\mathcal{A}/\mathcal{A}$ & 2061 (20\%)&6516 (20\%)&7456 (21\%)&4280 (42\%)&14341 (43\%)&15574 (44\%)&6618 (65\%)&22412 (68\%)&23930 (69\%)\\
                $\Delta{\dot{f}/\dot{f}}$&4744 (47\%)&8666 (28\%)&9091 (27\%)&6342 (63\%)&12257 (40\%)&12769 (37\%)&8611 (85\%)&19041 (61\%)&19772 (58\%)\\
                $\Delta{\cos{\iota}}$&3375 (33\%)&11062 (34\%)&12218 (35\%)&5015 (49\%)&16755 (51\%)&18026 (51\%)&6946 (68\%)&23349 (71\%)&24920 (71\%)\\
                $\Delta{\psi}$&2635 (26\%)&9282 (28\%)&10206 (29\%)&4063 (39\%)&13725 (42\%)&14841 (42\%)&5648 (56\%)&18948 (57\%)&20255 (58\%)\\
                $\Delta{\mathcal{M}/\mathcal{M}}$&5898 (58\%)&11169 (36\%)&11704 (34\%)&7698 (76\%)&15800 (51\%)&16400 (48\%)&9443 (94\%)&23734 (76\%)&24650 (72\%)\\
                $\Delta{\mathcal{D}/\mathcal{D}}$&1769 (17\%)&3931 (12\%)&4599 (13\%)&3446 (34\%)&7645 (24\%)&8317 (24\%)&6000 (59\%)&14829 (47\%)&21570 (63\%)\\
            \hline
    	\end{tabular*}}
    \end{threeparttable}    
\end{table*}

For Model 1 (Model 2), Table \ref{tab:cum_mdo} shows the number of DWDs with $\Delta{\mathcal{M}/\mathcal{M}}$, $\Delta{\mathcal{D}}/\mathcal{D}$, and $\Delta{\mathrm{\Omega}}$ of less than 0.001, 0.1, and 1 $\rm{deg^2}$ for TianQin, LISA, and their joint observation, respectively. Overall, the joint observation can increase the numbers for LISA by approximately 20\%. These sources are likely to be measured with the highest precision by the space-based detectors and can facilitate the subsequent study of the initial mass function and the structure of our Galaxy.

\begin{table}
    \caption{The expected numbers of Galactic DWDs detectable by TianQin, LISA, and their joint observation with $\rho \geq 7(5)$ over various mission durations. The results for $\rho \geq 5$ are given for comparison with other works. }
    \label{table:number_GW}
    \begin{tabular}{cccc}
    \hline
    Detector&1yr&2yr&5yr\\
  \hline
    \multicolumn{4}{|c|}{Model 1}\\
    TianQin& 2183 (3133)&3165 (4582)&4969 (7075)\\
    LISA& 5072 (6436)&6231 (8042)&16960 (23889)\\ 
    TianQin+LISA&5482 (7012)&6923 (9144)&18087 (25548)\\
    \hline
    \multicolumn{4}{|c|}{Model 2}\\
    TianQin& 4413 (6477)&6541 (9372)&10180 (14718)\\
    LISA& 9383 (12242)&12341 (15993)&32815 (44809)\\ 
    TianQin+LISA&10380 (13765)&13888 (18346)&34917 (47832)\\
    \hline
    \end{tabular}
\end{table}

\begin{table}
 \caption{The number of DWDs with $\Delta{\mathcal{M}/\mathcal{M}}$, $\Delta{\mathcal{D}}/\mathcal{D}$, and $\Delta{\mathrm{\Omega}}$ of less than 0.001, 0.1, and 1 $\rm{deg^2}$ for TianQin, LISA and their joint observation in Model 1 (Model 2).} 
 \label{tab:cum_mdo}
 \begin{tabular*}{\columnwidth}{@{\hspace*{5pt}}c@{\hspace*{20pt}}c@{\hspace*{25pt}}c@{\hspace*{25pt}}c@{\hspace*{25pt}}}

  \hline
  & TQ & LISA & TQ+LISA\\[2pt]
  \hline
  $\Delta{\mathcal{M}/\mathcal{M}}<0.001$&216 (543) &283 (727) &391 (907)\\[2pt]

  $\Delta{\mathcal{D}/\mathcal{D}}<0.1$ & 915 (1769)&1913 (3931) &2243 (4599)\\[2pt]

  $\Delta{\mathrm{\Omega}}<1 \rm{deg^2}$ &926 (1889) & 1896 (4010)&2275 (4722)\\[2pt]
  \hline
 \end{tabular*}
\end{table}

\section{Joint observation}\label{sec:joint}

In the previous two sections, we examined separately the capabilities of Gaia and VRO in optical observations and the capabilities of TianQin and LISA in GW detections for the DWDs in our Galaxy. In this section, we examine the properties of the DWDs detectable by both GW and EM observations. 

In Table \ref{table:joint_EM_GW}, we show the numbers of the DWDs with SNRs greater than 7 in Model 1 and Model 2 during the 5-year and 4-year nominal mission durations for TianQin and LISA, respectively. In addition, we count the numbers of eclipsing DWDs, constrained by the criteria such as sky area, apparent magnitude, and chi-square test, within Model 1 and Model 2 during the 5-year and 10-year nominal mission lifetimes of Gaia and VRO, respectively. The numbers of DWDs detectable by either EM or GW observations in Model 2 are about 2-3 times larger than in Model 1. The detection of DWDs by GW is influenced by a number of factors, including the source distance, spatial position, mass, orbital period, and so forth. The main factor contributing to the differences in the numbers of GW detections between Model 1 and Model 2 is the orbital period distribution of simulated DWDs. For example, in Model 2, the thin disk of our Galaxy follows a constant star formation history over a period of 10 billion years, resulting in more DWDs with frequencies in the millihertz range. On the other hand, we find that the main factor causing the difference in the numbers of EM detections between Model 1 and Model 2 is the effective surface temperature of the companion stars. We can see that the joint observation of TianQin and Gaia identifies 2 (Model 1) and 9 (Model 2) eclipsing DWDs. Similarly, the joint observation of TianQin and VRO identifies 7 (Model 1) and 11 (Model 2) eclipsing DWDs. In addition, the joint observation of LISA and Gaia identifies 4 (Model 1) and 29 (Model 2) eclipsing DWDs, while the joint observation of LISA and VRO identifies 15 (Model 1) and 47 (Model 2) eclipsing DWDs. 

\begin{figure}
    \centering
    
    \subfigure[]{
    \label{fig.dis.d}
    \includegraphics[width=0.38\textwidth]{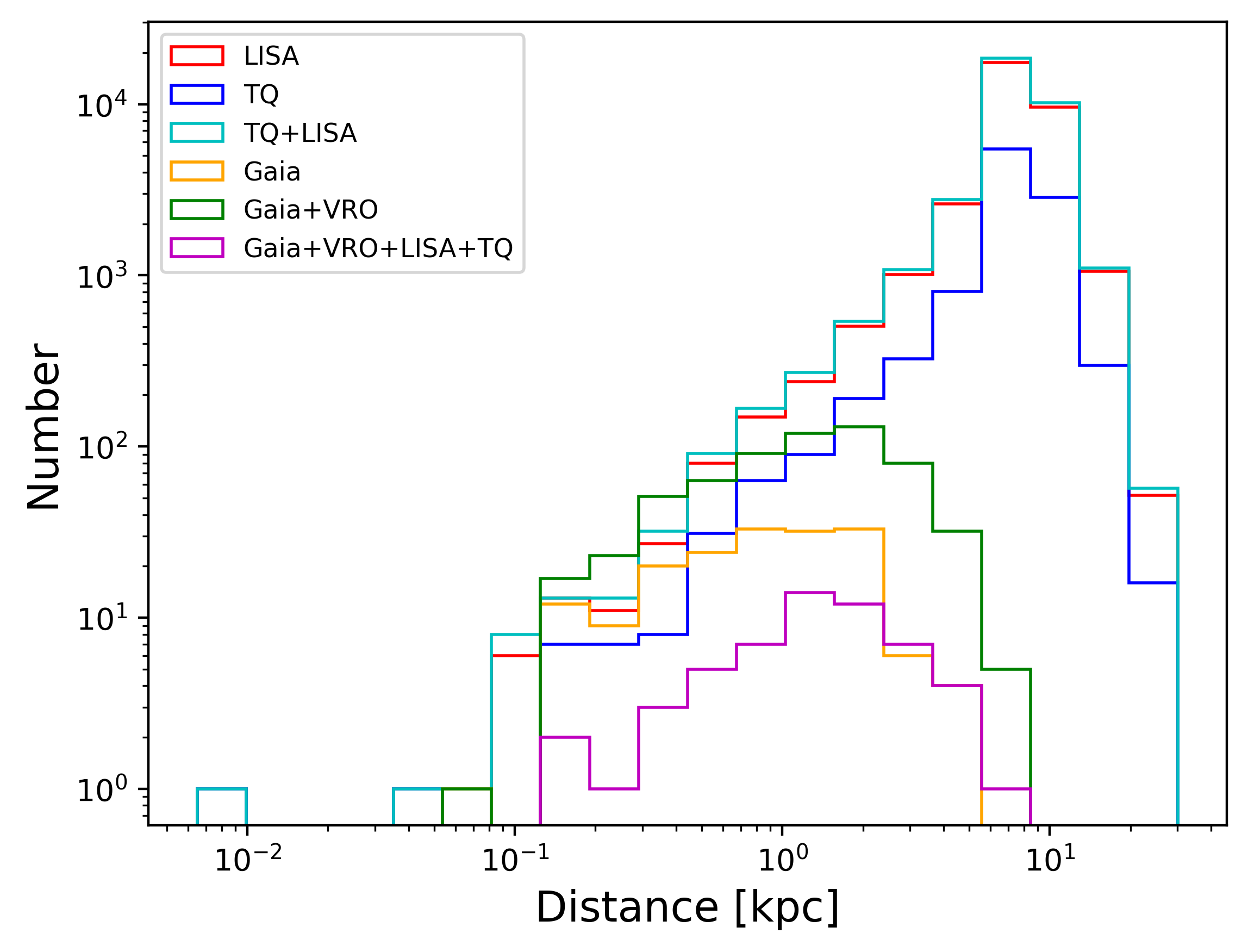}
    }\\[5pt]
    
    \subfigure[]{
    \label{fig.dis.p}
    \includegraphics[width=0.4\textwidth]{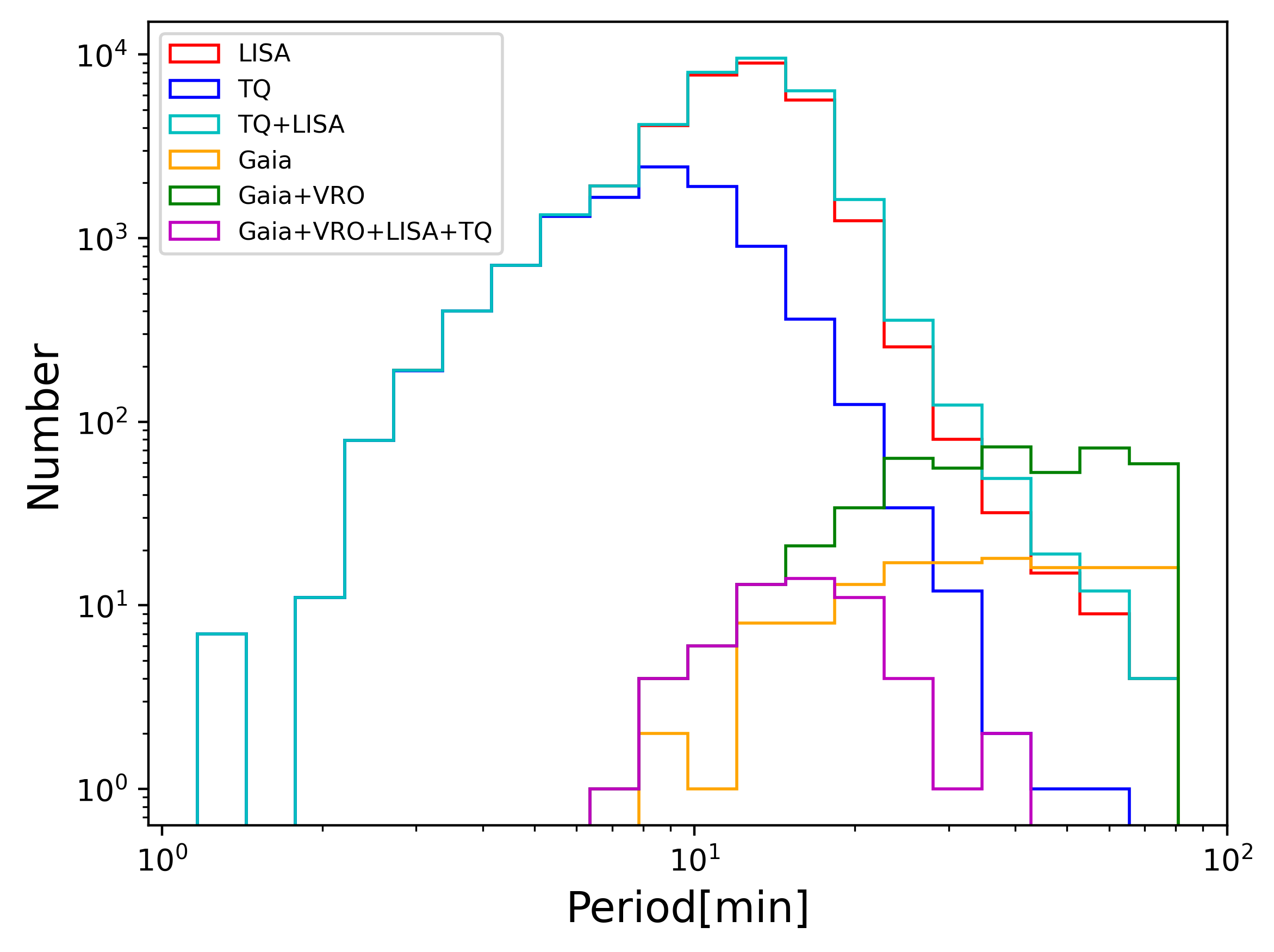}
    }\\[5pt]
    
    \subfigure[]{
    \label{fig.dis.mc}
    \includegraphics[width=0.38\textwidth]{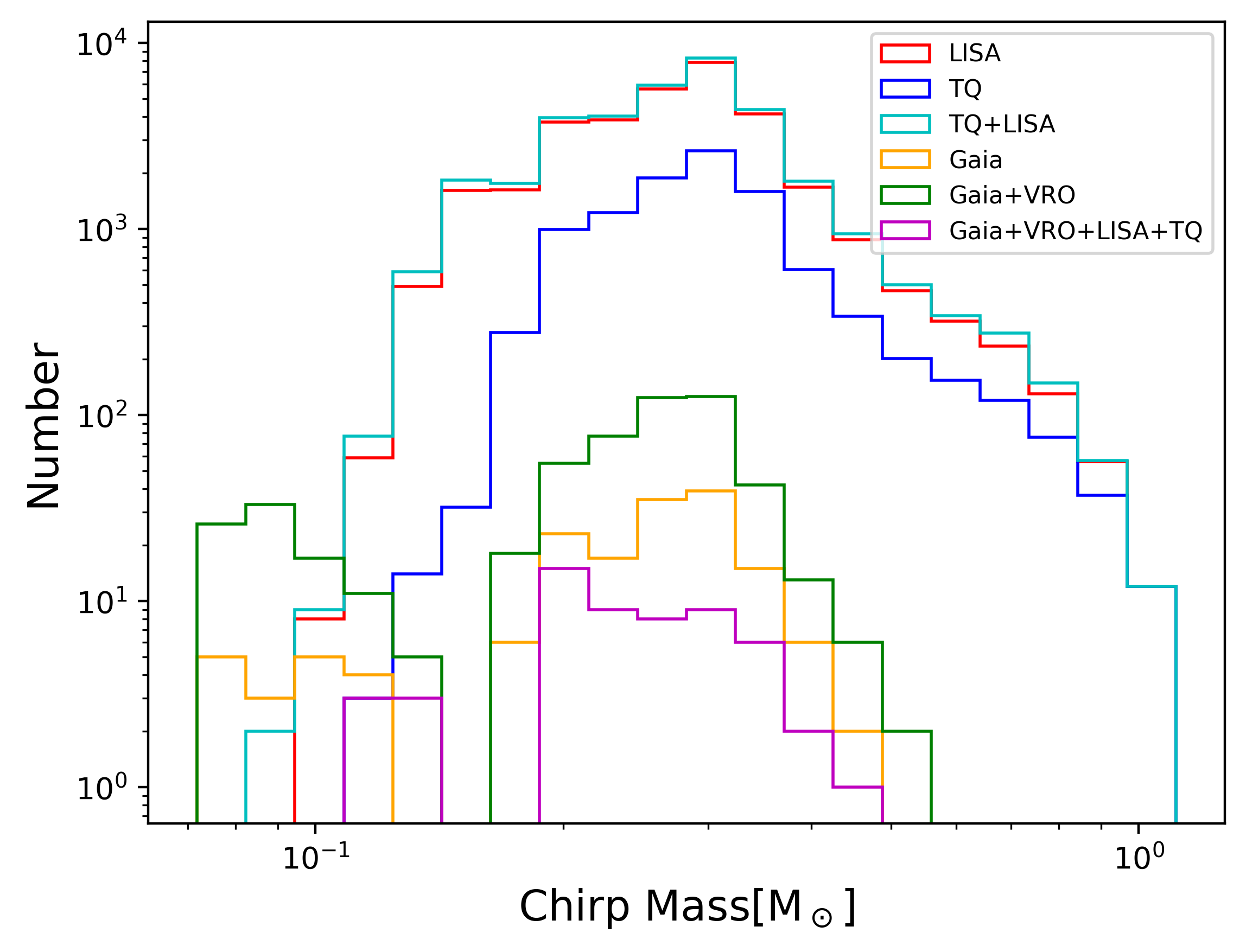}
    }
    
    \caption{Histograms of source distance, orbital period, and chirp mass for different detector combinations.}
    \label{fig:joint_distribution_D_M_P}
\end{figure}

Figure \ref{fig:joint_distribution_D_M_P} shows the distributions of distance, orbital period, and chirp mass of the DWDs detectable by EM and GW observations. From Figure \ref{fig.dis.d}, we can see that 95\% of the Galactic DWDs detectable by GW detectors are within 12 kpc, while 95\% of the eclipsing DWDs observed by optical telescopes are within 3.87 kpc. Their spacial distributions can be found in Figure \ref{fig:Galactic_x_z}. It is evident that the GW detectors can certainly make a more thorough survey of the DWDs in our Galaxy.

In Figure \ref{fig.dis.p}, we find that 97\% of the DWDs detectable by GW detectors have orbital periods shorter than 20 minutes. This is because the most sensitive frequency range of space-borne GW detectors is $10^{-3}-10^{-2}$ Hz, which corresponds to orbital periods of $3-33$ minutes. However, about only 2\% of the DWDs in the entire population of our simulation have orbital periods shorter than 33 minutes. 
In contrast, 90\% of eclipsing DWDs observed by optical telescopes have periods longer than 20 minutes. This partially explains why the number of DWDs jointly detectable by both GW detectors and optical telescopes is quite limited, and 90\% of the DWDs have orbital periods between 8 and 25 minutes.

Figure \ref{fig.dis.mc} shows the chirp mass distributions, which peak around 0.3 $M_{\sun}$, with 76\% of the DWDs distributed between 0.2 $M_{\sun}$ and 0.4 $M_{\sun}$ for the GW detectors. 
In terms of chemical composition, these DWDs are mainly He-CO and He-He DWDs, which is consistent with the case for the entire DWD population as shown in Table \ref{table:number_DWDs}. 
It is also clear that the GW detectors can find heavier DWDs with masses greater than 0.5 $M_{\sun}$ than the optical telescopes. 
On the other hand, WDs with a smaller mass typically have a larger radius, resulting in a larger radiating surface area. However, our analysis shows that compared to the radius, the surface temperature has a more significant effect on the optical detection of DWDs, which is actually a manifestation of the Stefan-Boltzmann law, $L = 4 \pi R^2\sigma T^4$, where $L$ is the radiation flux, $R$ is the radius, and $T$ is the surface temperature of the WD.

\begin{table}
    \caption{Number of DWDs from Model 1 (Model 2) that are detectable by TianQin, LISA, Gaia, VRO individually (diagonal elements), and their joint observation (off-diagonal elements).}
    \label{table:joint_EM_GW}
    \begin{tabular}{ccccc}
    \hline
    &TianQin&LISA&Gaia&VRO\\
  \hline
    TianQin&4969 (10180)&4969 (10180)&2 (9)&7 (11)\\
    LISA&4969 (10180)&16960 (32815)&4 (29)&15 (47)\\
    Gaia& 2 (9)&4 (29)&67 (174)&51 (116)\\ 
    VRO&7 (11)&15 (47)&51 (116)&273 (554)\\
    \hline
    \end{tabular}
\end{table}

\section{Summary and Conclusions}\label{sec:sum}

In this work, we simulated two populations of DWDs in our Galaxy with different star formation histories (Model 1 and Model 2) using the binary population synthesis method. Model 1 assumes a burst of star formation 13.7 Gyr ago, while Model 2 assumes a constant star formation rate for the thin disk and a burst of star formation for the thick disk and bulge.
We predicted the numbers of DWDs detectable by TianQin, LISA, Gaia, and VRO individually and jointly in these two samples. We found that TianQin can detect 4969 (10180) DWDs with SNRs greater than 7 within a 5-year nominal mission duration in Model 1 (Model 2), while LISA can detect 16960 (32815) DWDs with SNRs greater than 7 within a 4-year nominal mission duration in Model 1 (Model 2). The number of DWDs detectable by TianQin and LISA in Model 2 is found to be twice that of in Model 1. The joint observation of TianQin and LISA can detect an additional 2,000 DWDs compared to LISA alone in Model 2. 
For EM observations, we found that Gaia and VRO can detect 67 (174) and 273 (554) compact eclipsing DWDs in Model 1 (Model 2)  with orbital period
less than 30 hours, respectively. 
The increased number in Model 2 is mainly due to the higher surface temperatures of the WDs generated in Model 2, making them more easily detectable by optical telescopes. 

For the space-borne detectors, we calculated the SNRs of DWDs for TianQin and LISA, individually and jointly. The latter can detect about $7\%$ more DWDs compared to the LISA alone. Furthermore, we analyzed the estimation errors of the parameters using the Fisher information matrix. 
Due to the sensitive frequencies of the GW detectors, almost all DWDs detectable by LISA and TianQin are in the range of $10^{-3}-10^{-2}$~Hz, corresponding to orbital periods of 3-30 mins. In contrast, the orbital periods of DWDs observed in EM observations are mainly centered around one hour. 

In the investigation of EM observations of eclipsing DWDs, we found that the $r$-band of VRO has superior performance compared to other bands, with a limiting magnitude of 24.7. The DWDs detectable by optical telescopes are predominantly distributed near the Galactic disk.  

In this work, we have focused solely on simulated observations from VRO and Gaia, but it is important to recognize that many other survey projects will also contribute to the detection of more DWDs.
Since VRO primarily observes the southern hemisphere, including surveys such as ZTF and other northern hemisphere facilities is essential to achieve broader sky coverage. 
For example, the Chinese Space Station Telescope (CSST)~\citep{2019ApJ...883..203G} with a $g$-band magnitude of 26.4 and a field of view of about 1.1 deg$^2$ is expected to identify more DWDs candidates during its primary survey mission. 
The current sample of known DWDs is obtained through data mining across a combination of photometric, spectral, and astrometric sky surveys, supplemented by dedicated follow-up observations. 
To detect more DWDs using EM methods in the future, a similar approach will be necessary—even in the VRO era—by first identifying DWD candidates from large-scale surveys and then conducting targeted follow-up observations using both ground-based and space-based telescopes. 

Furthermore, much works remain to be done in the near future. 
For instance, our current simulation only considers the detached DWDs, and we plan to extend our study to include semi-detached DWDs.
Additionally, direct comparisons between our population synthesis results and both existing and forthcoming EM surveys of DWDs, such as those from the Zwicky Transient Facility \citep[ZTF;][]{2023ApJS..264...39R}, Gaia, the Wide Field Survey Telescope \citep[WFST;][]{Wang23}, and the Multi-channel Photometric Survey Telescope \citep[Mephisto;][]{Yuan20} will be crucial in refining our binary population synthesis models and improving their accuracy. 

\section*{Acknowledgements}

We thank Valeriya Korol, Shenghua Yu and Xiang-Dong Li for helpful discussions. 
HMJ is grateful for the hospitality of Nanjing University, where part of this work was done during his visit. 
YW gratefully acknowledges support from the National Key Research and Development Program of China (No. 2023YFC2206702 and No. 2022YFC2205201), the National Natural Science Foundation of China (NSFC) under Grants No. 11973024, Major Science and Technology Program of Xinjiang Uygur Autonomous Region (No. 2022A03013-4), and Guangdong Major Project of Basic and Applied Basic Research (No. 2019B030302001). YS acknowledges support from the National Key Research and Development Program of China (No. 2021YFA0718500 and No. 2023YFA1607902). BM acknowledges support from NSFC (No. 12073092 and No. 12103097).
We acknowledge the High Performance Computing Platform at Huazhong University of Science and Technology for providing computational resources. 
The authors thank the anonymous referee for helpful comments and suggestions that significantly improved our manuscript.


\section*{Data Availability}


The simulation data underlying this article will be shared on reasonable request to the corresponding authors.



\bibliographystyle{mnras}
\bibliography{example} 






\bsp	
\label{lastpage}
\end{document}